\documentclass[utf8]{aa}
\usepackage{graphicx}
\usepackage[english]{babel}
\usepackage{multirow}
\usepackage{latexsym}
\usepackage{amssymb}
\usepackage{relsize}
\usepackage{txfonts}
\usepackage{array} 
\usepackage{tablefootnote}
\bibpunct{(}{)}{,}{a}{}{,}
\usepackage[dvipsnames]{xcolor}
\usepackage{ulem}
\usepackage{amsmath}
\usepackage{hyperref}
\usepackage{verbatim}
\usepackage{upgreek}
\usepackage{bm}
\hypersetup{
    colorlinks=true,
    allcolors=blue
}
\usepackage[switch]{lineno} 

\newcommand{\Phieff}{\Phi_{\mathrm{eff}}}
\newcommand{\PhiG}{\Phi_{\mathrm{G}}}
\newcommand{\PhiC}{\Phi_{\mathrm{C}}}

\newcommand{\vect}{\vec}

\renewcommand{\dfrac}[2]{\frac{\mathrm{d} #1}{\mathrm{d} #2}}
\newcommand{\ddfrac}[2]{\frac{\mathrm{d}^2 #1}{\mathrm{d} #2^2}}

\newcommand{\grad}{\vect{\nabla}}

\newcommand{\lapl}{\Delta}

\renewcommand{\l}{\ell}

\newcommand{\rz}{r_{\zeta}}
\newcommand{\rt}{r_{\theta}}

\newcommand{\dz}{\partial_{\zeta}}

\newcommand{\dt}{\partial_{\theta}}

\newcommand{\dzt}{\partial_{\zeta\theta}^2}
\newcommand{\dtt}{\partial_{\theta\theta}^2}

\newcommand{\cott}{\cot\theta}

\newcommand{\OmegaK}{\Omega_{\mathrm{K}}}

\newcommand{\PhiGl}{\Phi_{\mathrm{G}}^{\l}}
\newcommand{\Ylm}{Y^m_{\ell}}

\newcommand{\rsph}{r_{\mathrm{sph}}}
\newcommand{\Rpol}{R_{\mathrm{pol}}}
\newcommand{\Req}{R_{\mathrm{eq}}}

\makeatletter
\DeclareRobustCommand\onedot{\futurelet\@let@token\@onedot}
\def\@onedot{\ifx\@let@token.\else.\null\fi\xspace}

\def\eg{\textit{e.g}\onedot} 
\def\ie{\textit{i.e}\onedot} 
\def\cf{\textit{c.f}\onedot}

\makeatother

\colorlet{mdtRed}{red!80!black}
\colorlet{mdtGray}{lightgray!70!gray}
\colorlet{mdtOrange}{orange!100!black}

\begin{document}
   \title{RUBIS: a simple tool for calculating the centrifugal deformation of stars and planets}

   \author{P. S. Houdayer \and D. R. Reese}

   \institute{
           LESIA, Observatoire de Paris, Université PSL, CNRS,
           Sorbonne Université, Univ. Paris Diderot, Sorbonne Paris Cité,
           5 place Jules Janssen, 92195 Meudon, France \\
           \email{pierre.houdayer@obspm.fr, daniel.reese@obspm.fr}
   }

   \date{Received 14 March 2023; Accepted 25 April 2023}

  \abstract
   {}
   {We present RUBIS (Rotation code Using Barotropy conservation over Isopotential Surfaces), a fully Python-based centrifugal deformation program available at \url{https://github.com/pierrehoudayer/RUBIS}. The code has been designed to calculate the centrifugal deformation of stars and planets resulting from a given cylindrical rotation profile, starting from a spherically symmetric non-rotating model.}
   {The underlying assumption in RUBIS is that the relationship between density and pressure is preserved during the deformation process.  This leads to many procedural simplifications.  For instance, RUBIS only needs to solve Poisson' equation, either in spheroidal or spherical coordinates depending on whether the 1D model has discontinuities or not.}
   {We present the benefits of using RUBIS to deform polytropic models and more complex barotropic structures, thus providing, to a certain extent, insights into baroclinic models. The resulting structures can be used for a wide range of applications, including the seismic study of models.  Finally, we illustrate how RUBIS is beneficial specifically in the analysis of Jupiter's gravitational moments, thanks to its ability to handle discontinuous models while retaining a high accuracy compared to current methods.}
   {}


   \maketitle
%

\section{Introduction}

Rotation is ubiquitous both in stars and planets, and has a profound effect on
both their structure and evolution.  For instance, recent interferometric
observations have shown to what extent stars can be affected by centrifugal
deformation and gravity darkening \citep[\eg][]{DomicianoDeSouza2003,
Monnier2007, Zhao2009, Che2011, Bouchaud2020}. Various studies have predicted \citep{Endal1976, Zahn1992, Maeder1998, Mathis2004} and shown \citep{Meynet2000, Palacios2003, Palacios2006, Amard2019} its impact on stellar
evolution and raised a number of open questions \citep[\eg][]{Deheuvels2012, Deheuvels2015, Benomar2015, Ouazzani2019}. The monograph \citet{Maeder2009} provides a comprehensive description of the impact of rotation on stellar evolution from a theoretical standpoint, while \citet{Aerts2019} provides a recent review that focuses on transport processes in such stars and the resulting open questions. Likewise, rotation plays an important role in gas giants such as Jupiter and Saturn.  Indeed, accurately taking into account centrifugal deformation proved to be critical when interpreting the gravitational moments of Jupiter measured by Juno in order to investigate the presence of a core \citep[\eg][]{Wahl2017} or to probe the wind gradient and differential rotation \citep[\eg][] {Iess2018, Guillot2018}. 
Furthermore, the recent detection of f-modes\footnote{\textit{Fundamental (f) modes}: oscillation modes with no radial nodes, thus with the radial order $n=0$.} \citep[\eg][]{Hedman2013} and g-modes \citep{Mankovich2021} in Saturn has sparked kronoseismic\footnote{\textit{Kronoseismology}: seismology of Saturn.} investigations into Saturn's core which required taking into account the effects of rotation on Saturn's structure and pulsations \citep{Fuller2014, Mankovich2019, Dewberry2021}.
Hence, there is a real need for numerical
tools able to calculate the structure of such stars and planets.

In the stellar domain, much progress has been made over the past years in
devising 2D stellar structure codes that fully take rotation into account.  For
instance, the Self-Consistent Field (SCF) method has been devised to calculate
the structure of stars with pre-imposed cylindrical rotation profiles
\citep{Jackson2005, MacGregor2007}.  It alternates between solving
Poisson's equation and the hydrostatic equilibrium, thereby iteratively
adjusting the distribution of matter in  the star.  Given that the rotation
profile is conservative, the structure of the star is barotropic, \ie\ lines of
constant pressure, density, temperature, and thus total
(gravitational+centrifugal) potential coincide.  Hence, solving the hydrostatic
equilibrium amounts to finding the lines of constant total potential and
redistributing the matter so that the density is constant along these lines.

A drawback with the SCF method is that the energy equation is only solved along
horizontal averages rather than locally.  To overcome this difficulty, the
Evolution STEllaire en Rotation (ESTER) code was developed
\citep{EspinosaLara2013, Rieutord2016}.  As a result of solving the energy
equation locally, the rotation profile (which is calculated along with the
stellar structure) is no longer conservative and the stellar structure is
baroclinic, \ie isodensity and isobars are now free to differ.  Currently, neither code is capable of carrying out stellar
evolution and instead calculate static models.  The composition within ESTER
models may nonetheless be adjusted to mimic the effects of stellar evolution.

A solution for bypassing the above limitation is to take stellar models from 1D
non-rotating stellar evolution codes and to subsequently introduce the effects
of centrifugal deformation. This is the strategy introduced in
\citet{Roxburgh2006}. Indeed, he uses the density profile from the 1D model,
applies it along a radial cut (at some given latitude), and then iteratively
reconstructs the distribution of matter in the rest of the star for a predefined
2D rotation profile.  From there, the pressure distribution and gravitational
field may be calculated.  In addition, provided an assumption is made on the
chemical composition, then it is possible to deduce the adiabatic exponent,
$\Gamma_1$, throughout the star thanks to the equation of state, followed by
variables such as the sound velocity and the Brunt-V{\"a}is{\"a}l{\"a}
frequency. Hence, a complete acoustic structure is obtained thus allowing the
calculation of pulsation modes \citep{Ouazzani2015}.

One of the limitations of this method is that the energy equation is not taken
to account and is thus not generally satisfied.  To overcome this difficulty,
one may apply this method to 1D models where the effects of rotation are already
taken into account, albeit in an approximate way.  For instance, models from
STAREVOL \citep{Palacios2003,Palacios2006}, the Geneva stellar evolution code
\citep{Eggenberger2008}, and CESTAM \citep{Marques2013} take into account the
horizontally averaged effects of rotation thanks to the formalism developed in
\citet{Zahn1992} and \citet{Maeder1998}.  \citet{Manchon2021} developed a method which
goes a step further since it calculates the centrifugal deformation for CESTAM
models, but then feeds the information from the 2D structure back into the 1D
formalism using the approach described in \citet{Mathis2004}.  Such a procedure
could then be applied at each time step when calculating the evolution of a
star, thus achieving a greater degree of realism.

In the present article, we wish to develop a method analogous to that of
\citet{Roxburgh2006} but which is simpler.  In particular, we wish to avoid
having to reuse the equation of state to recalculate the $\Gamma_1$ profile. 
The resulting program, \href{https://github.com/pierrehoudayer/RUBIS}{RUBIS} (Rotation code Using Barotropy conservation over Isopotential Surfaces), achieves this by
preserving the relation between density and pressure
rather than density and radius when going from the 1D to the 2D structure as
will be described below.  Hence, the equation of state will automatically be
satisfied throughout the model given that thermodynamic quantities are simply
carried over from the 1D case.  With such an approach, deforming a 1D polytropic
structure, for instance, will lead to a 2D polytropic structure unlike what
would happen with the approach in \citet{Roxburgh2006}.  Finally, we wish to
make this approach applicable both to stars and planets which may include
density discontinuities.  This presence of a discontinuity will lead to
a different strategy when solving Poisson's equation as will be explained
below.

As was the case with the approach developed in \citet{Roxburgh2006}, RUBIS does not take into account the energy conservation equation. 
Hence, the models it deforms will only be suitable for adiabatic pulsation
calculations.  For full non-adiabatic calculations, one should instead use
models such as those from the ESTER code which solves the hydrostatic structure
and energy equation in a full 2D context.

The article organised as follows: Section~\ref{sect:RUBIS} begins by explaining how our code, RUBIS, works, and Section~\ref{sect:specificities} emphasises the specific features that differentiate it from existing programs. Section~\ref{sect:tests} will be devoted to carrying out numerical tests, in particular to comparisons with existing programs and to establishing the scope in which the assumptions adopted are valid. Section~\ref{sect:conclusion} will be dedicated to our conclusion and more broadly to our perspectives on the future use of RUBIS.

\section{Description of RUBIS \label{sect:RUBIS}}

As stated in the introduction, the starting point of the method is to assume that
the relation between
$\rho$ and $P$ is preserved when going from the non-rotating to the rotating models.
This can be justified in one of two ways: either there exists some intrinsic relation
between $\rho$ and $P$, \eg\ the polytropic relation, or we assume the thermodynamic 
structure of the non-rotating model is in some sense a good approximation to that
of the rotating model, an assumption we will discuss in Sections~\ref{sect:specificities} \& \ref{sect:tests}. We will now derive the direct implication of this assumption, which is also RUBIS' key property.

The hydrostatic equilibrium of a rotating, self-gravitating object (with a conservative rotation
profile) is described by the following equation:
\begin{equation}
\grad P = -\rho \grad \Phieff,
\label{eq:hydrostatic}
\end{equation}
where $P$ is the pressure, $\rho$ the density, and $\Phieff = \PhiG + \PhiC$ the
total potential, $\PhiG$ being the gravitational potential and $\PhiC$ the
centrifugal potential.  These potentials satisfy the following equations:
\begin{eqnarray}
\lapl \Phi_{\mathrm{G}} &=& 4 \pi \mathcal{G} \rho, \\
\label{eq:def_PhiC}
\Phi_{\mathrm{C}} &=& -\int_0^{s} \Omega^2(s') s' ds',
\end{eqnarray}
where $\mathcal{G}$ is the gravitational constant, $\Omega(s)$ the rotation
profile, and $s$ the distance to the rotation axis.  We recall that conservative
rotation profiles are cylindrical, \ie, they only depend on the distance to the
rotation axis.  It is this property that allows the centrifugal force to derive
from a potential and leads to a barotropic structure for the object, as can be
seen by taking the curl of Eq.~(\ref{eq:hydrostatic}) divided by the density: $\grad \rho \times \grad P = \mathbf{0}$.

Let us introduce a coordinate system $(\zeta,\theta,\varphi)$ such that $\zeta$ is constant along isopotential lines and where $\theta$ and $\varphi$ denotes the usual polar and azimuthal angles. Given that, at this stage, $\zeta$ is a dimensionless variable used to label the isopotential lines rather than being a physical coordinate, there are no requirements on what values it takes as it long as it varies smoothly and monotonically from the centre of the star or planet to the surface.
Accordingly, given the barotropic structure of the deformed object, quantities such as $P$ and $\rho$ only depend on $\zeta$. 
Hence, projected on the natural basis, one can show that Eq.~(\ref{eq:hydrostatic}) reduces to:
\begin{equation}
\dfrac{P}{\zeta} = -\rho \dfrac{\Phieff}{\zeta},
\end{equation}
the other components being zero. 
This equation can be compared with its
non-rotating equivalent:
\begin{equation}
\dfrac{P_{\mathrm{sph}}}{\rsph} = -\rho_{\mathrm{sph}}
                     \dfrac{\Phi_{\mathrm{sph}}}{\rsph},
\end{equation}
where $\Phi_{\mathrm{sph}}$ reduces to the gravitational potential, and where
the notation $\rsph$ has been introduced to avoid confusion between
$r$ in the rotating and non-rotating models.  It is possible to chose the values of $\zeta$
in such a way that $P(\zeta) \equiv P_{\mathrm{sph}}(\rsph)$ given that $P$ is a monotonic
function of $\zeta$.  As a result,  $dP/d\zeta = dP_{\mathrm{sph}}/d\rsph$ immediately
follows.  Furthermore, preserving the relation between $\rho$ and $P$ when going from
the non-rotating and to the rotating model also leads to $\rho(\zeta) =
\rho_{\mathrm{sph}}(\rsph)$, and this regardless of whether or not $\rho$
varies monotonically with $\zeta$. The hydrostatic
equations then show that $d\Phieff/d\zeta = d\Phi_{\mathrm{sph}}/d\rsph$.  In other
words:
\begin{equation}
\Phieff(\zeta) = \Phi_{\mathrm{sph}}(\rsph) + \mathrm{cnst}.
\label{eq:potential_1D}
\end{equation}
The constant that appears in this equation may in fact be deduced by applying
the above equation at the object's centre, \ie\ $\rsph=\zeta=0$ (assuming the
gravitational components of both potentials match a vacuum potential outside the
deformed object).

\begin{figure*}[htbp]
\begin{center}
\includegraphics[width=\textwidth]{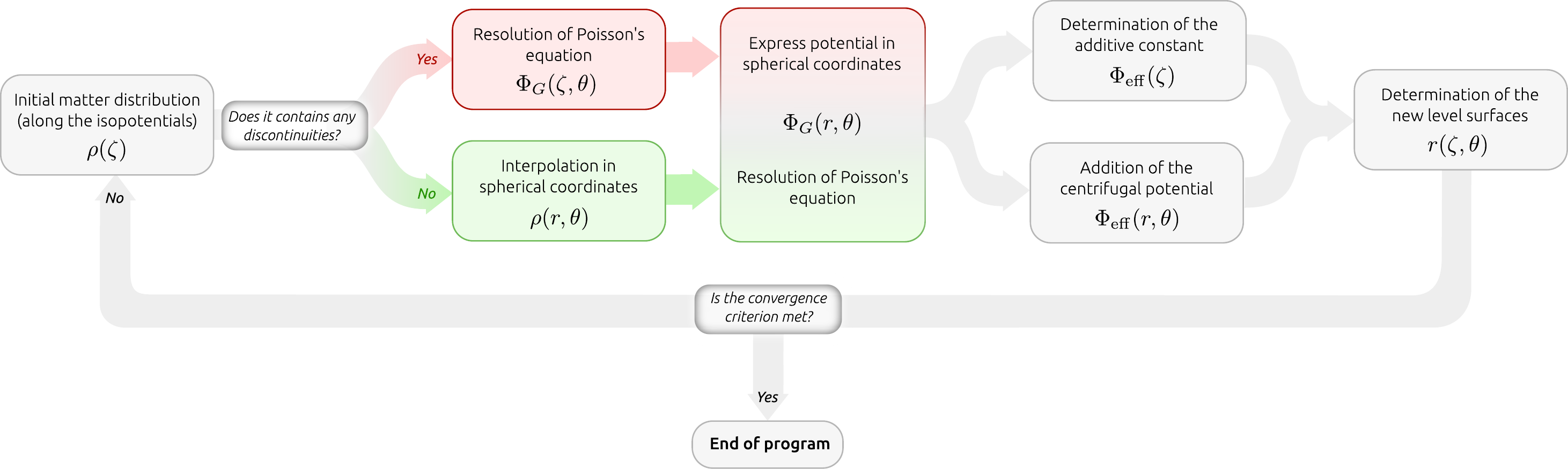}
\end{center}
\caption{Flowchart illustrating how RUBIS works. Each step shows the quantity that is obtained and in terms of which variable it is obtained.
\label{fig:schema_code}}
\end{figure*}

The above observations lead us to outlining the following iterative approach
which is a simplified version of the SCF algorithm
\citep{Jackson2005, MacGregor2007}:
\begin{enumerate}
\item Find the gravitational potential based on the matter distribution (Sect.~\ref{par:gravitational_potential})
\item Add the centrifugal potential to the gravitational one (Sect.~\ref{par:centrifugal_potential}).
\item Find the level surfaces, \ie\ the lines of constant total potential, and redistribute the matter on these surfaces (Sect.~\ref{par:level_surfaces}).
\item Return to step 1 and iterate till convergence (Sect.~\ref{par:convergence}).      
\end{enumerate}
Figure~\ref{fig:schema_code} illustrates schematically this algorithm. These steps
are described in more detail in what follows.

\subsection{Finding the gravitational potential \label{par:gravitational_potential}}

There are two different ways of calculating the gravitational potential:
\begin{itemize}
\item Interpolating the matter distribution onto the spherical coordinate system and solve Poisson's equation after having projected it onto the spherical harmonic basis paragraph~\ref{par:Poisson_spherical}).
\item Solving Poisson's equation using directly the spheroidal coordinate system (paragraph~\ref{par:Poisson_spheroidal}).
\end{itemize}

\subsubsection{Using spherical coordinates \label{par:Poisson_spherical}}

Although the first approach may sound more complicated due to the extra
interpolation step, it is in fact more efficient.  Indeed, once the matter
distribution is expressed using spherical coordinates, Poisson's equation
becomes separable with respect to the spherical harmonic basis thus allowing
it to be solved efficiently.

The interpolation of the density profile onto the spherical coordinate system is carried out using
cubic splines along each latitude thanks to the SciPy sub-package \href{https://docs.scipy.org/doc/scipy/reference/interpolate.html}{\texttt{scipy.interpolate}}. Afterwards,
the density profile is decomposed using the spherical harmonic basis as follows:
\begin{equation}
\rho(r,\theta,\varphi) = \sum_{\l=0}^{\infty} \sum_{m=-\l}^{\ell} \rho_\ell^m(r) \Ylm(\theta,\varphi),
\label{eq:harmonic_decomposition}
\end{equation}
where
\begin{equation}
\rho_\ell^m(r) = \iint_{4\pi} \rho(r,\theta,\varphi) \left[ \Ylm(\theta,\varphi) \right]^*
            \sin\theta d\theta d\varphi,
\end{equation}
and where $\l$ represents the harmonic degree, $m$ the azimuthal order, $\Ylm$
the corresponding spherical harmonic, and $(\cdot)^*$ the complex conjugate.  
Given that the deformed object is axisymmetric, only the $m=0$ components of $\rho$ and
$\PhiG$ are non-zero.  Hence, in what follows, we will be assume $m=0$ and will
drop the $m$ index. On a practical level, the manipulation of harmonic series, whether for decomposition or projection, is implemented in RUBIS using the appropriate \href{https://docs.scipy.org/doc/scipy/reference/special.html}{\texttt{scipy.special}} routines.

Poisson's equation is subsequently projected onto the spherical harmonic basis,
thus leading to
\begin{equation}
\ddfrac{\PhiGl}{r} + \frac{2}{r} \dfrac{\PhiGl}{r} - \frac{\l(\l+1)}{r^2} \PhiGl
= 4 \pi \mathcal{G} \rho_\ell,
\label{eq:Poisson_harmonic}
\end{equation}
for each spherical harmonic.  Here, $\PhiGl$ represents the projection of
$\PhiG$ onto the harmonic basis.  Equation~(\ref{eq:Poisson_harmonic}) can be
solved analytically using integrals:
\begin{equation}
\PhiGl = -\frac{4\pi\mathcal{G}}{2\l+1} \left[\int_0^r \rho_\ell(s) \frac{s^{\l+2}}{r^{\l+1}} ds
       + \int_r^R \rho_\ell(s) \frac{r^{\l}}{s^{\l-1}} ds\right],
\end{equation}
where $R$ is the stellar radius.  However, when $\l$ becomes large, the above
formula can lead to poor numerical results.  Therefore, we prefer to solve
Eq.~(\ref{eq:Poisson_harmonic}) numerically by first casting it into a first
order system of two differential equations, discretising it using the
finite-difference approach described in \citet{Reese2013b}, and solving the
system with an efficient band matrix factorisation using the appropriate
\href{https://netlib.org/lapack/}{Lapack} wrapper available in \href{https://docs.scipy.org/doc/scipy/reference/linalg.lapack.html}{\texttt{scipy.linalg.lapack}}. This requires including boundary conditions which ensure the continuity of $\PhiGl$ and its derivative on the object's surface:
\begin{align}
    \label{eq:phi_continuity_bc}
    \Phi_G^{\ell, \mathrm{in}}(R) &= \Phi_G^{\ell, \mathrm{out}}(R) \\
    \label{eq:dphi_continuity_bc}
    \left.\frac{d\Phi_G^{\ell, \mathrm{in}}}{dr}\right|_R &= \left.\frac{d\Phi_G^{\ell, \mathrm{out}}}{dr}\right|_R
\end{align}
In addition, we also apply regularity conditions in the centre and at infinity:
\begin{align}
    \label{eq:regularity_centre}
    \Phi_G^{\ell}(r) & \underset{r \rightarrow 0}{\propto} r^{\ell}  \\
    \label{eq:regularity_infinity}
    \Phi_G^{\ell}(r) & \underset{r \rightarrow \infty}{\propto} r^{-(\ell+1)}
\end{align}
Combining the latter equation with Eqs.~\eqref{eq:phi_continuity_bc} and~\eqref{eq:dphi_continuity_bc} then leads to the following condition on the object's surface \citep[\eg][]{Ledoux1958}:
\begin{equation}
\dfrac{\PhiGl}{r}(R) + \frac{\l+1}{R} \PhiGl(R) = 0
\label{eq:Poisson_boundary_condition}
\end{equation}

Equations~\eqref{eq:Poisson_harmonic} together with the conditions~\eqref{eq:regularity_centre} and~\eqref{eq:Poisson_boundary_condition} are solved up to a certain degree, $L$, which is fixed by the user at the beginning of the procedure. Once the $\PhiGl(r)$ functions have been obtained, the potential is then deduced in
terms of the $(r,\theta)$ coordinates using:
\begin{equation}
\PhiG(r,\theta) = \sum_{\l=0}^{\infty}\PhiGl(r) Y_\ell^0(\theta).
\end{equation}

\subsubsection{Using spheroidal coordinates \label{par:Poisson_spheroidal}}

The above approach will not function correctly if the density profile is
discontinuous. Indeed, a discontinuity in the density profile will follow a
level surface due to the object's barotropic structure.  Hence, it will intersect
spherical surfaces thus causing the density profile to be discontinuous as a
function of latitude for certain values of $r$.  This will lead to poor numerical
results when the density profile is projected onto the spherical harmonic basis
and may stop the algorithm from converging.

Accordingly, Poisson's equation must be solved directly in the spheroidal
coordinate system, $(\zeta,\theta,\varphi)$.  Hence, the following harmonic
decomposition will be used instead:
\begin{equation}
\PhiG(\zeta,\theta) = \sum_{\l=0}^{\infty} \PhiGl(\zeta) Y_\ell^0(\theta),
\label{eq:harmonic_decomposition_spheroidal}
\end{equation}
If we treat the radial distance, $r$, as a function of $\zeta$ and $\theta$,
then thanks to tensor analysis (see Appendix~\ref{app:Poisson}), we obtain the following explicit expression for
Poisson's equation:
\begin{equation}
\dz\left(\frac{r^2+\rt^2}{\rz}\dz\PhiG\right)
- 2\rt\dzt \PhiG - \lapl_\mathcal{S}r\, \dz \PhiG
+\rz \lapl_\mathcal{S} \PhiG
= 4 \pi \mathcal{G} r^2\rz \rho,
\label{eq:Poisson_spheroidal}
\end{equation}
where:
\begin{equation}
\lapl_\mathcal{S} = \dtt + \cott \dt
\end{equation}
and where $r_{\zeta} = \dz r$, $r_{\theta} = \dt r$, $r_{\zeta\theta}=\dzt r$, etc.
We note that due to symmetry around the rotation axis, derivatives with respect
to $\varphi$ vanished.

This equation is then projected onto the spherical harmonic basis, discretised
in the radial direction using finite-differences, and solved.  Given its
expression, Eq.~\eqref{eq:Poisson_spheroidal} is not separable on the spherical
harmonic basis. Hence, the equations for the different $\PhiGl$ are coupled: 
\begin{equation}
    \label{eq:Poisson_spheroidal_proj}
    \sum_{\ell' = 0}^L \dz\left(\mathcal{P}_{\zeta \zeta}^{\ell\ell'} \dz \PhiG^{\ell'}\right) - \mathcal{P}_{\zeta \theta}^{\ell\ell'}\dz \PhiG^{\ell'} - \mathcal{P}_{\theta \theta}^{\ell\ell'} \PhiG^{\ell'} = 4 \pi \mathcal{G} {(r^2\rz)}_\ell~ \rho
\end{equation}
which results in the appearance of coupling integrals, $\mathcal{P}^{\ell \ell'}_{\cdot\cdot}$, as well as the harmonic decomposition of $r^2\rz$ denoted ${(r^2\rz)}_\ell$  (please refer to Appendix~\ref{app:Poisson} for an explicit expression of the terms appearing in Eq.~\ref{eq:Poisson_spheroidal_proj}). These equations must then be solved simultaneously. 
However, the use of finite differences means
that only adjacent values of $\zeta$ are coupled.  Hence, grouping together the
unknowns, $\PhiGl(\zeta_i)$, according to $\zeta$ values leads to a band matrix
(although of much larger dimensions than when solving
Eq.~(\ref{eq:Poisson_harmonic})) which can be filled efficiently using the \href{https://docs.scipy.org/doc/scipy/reference/sparse.html}{\texttt{scipy.sparse}} package and once more solved with the Lapack routine. 

In order to ensure that the potential matches a vacuum potential outside the
deformed object, a second domain is added with the object's surface as an inner boundary and
a sphere as the outer boundary.  Poisson's equation is then enforced on this domain subject to
interface conditions on the inner boundary in order to ensure the continuity of
the gravitational potential and its gradient which, in terms of the spheroidal coordinates, results in preserving $\rz^{-1}\dz \PhiG$ in addition to $\PhiG$ through the surface. In the case of internal density discontinuities, these two conditions (derived in Appendix~\ref{app:Poisson}, \cf Eqs.~\ref{eq:interface_phi} and \ref{eq:interface_dphi}) must be added at each of the domain interfaces. Finally, conditions analogous to Eqs.~\eqref{eq:regularity_centre} and~(\ref{eq:Poisson_boundary_condition}) are applied in the centre and on the outer spherical boundary.

Once more, $\PhiG(\zeta,\theta)$ can subsequently be deduced from the
$\PhiGl(\zeta)$ using Eq.~(\ref{eq:harmonic_decomposition_spheroidal}).  It should be noted that the latter equation implicitly also gives us the gravitational potential as a function of $r$ and $\theta$ by using the relation $r(\zeta, \theta)$ since $\PhiG(r(\zeta, \theta),\theta)~=~\PhiG(\zeta,\theta)$.

\subsection{Adding the centrifugal potential \label{par:centrifugal_potential}}

Whether solving Poisson's equation in spherical (cf. paragraph \ref{par:Poisson_spherical}) or spheroidal (cf. paragraph \ref{par:Poisson_spheroidal}) coordinates, we derived the gravitational potential $\PhiG(r,\theta)$ at this point. The total potential at any point in the structure is therefore determined by simply adding the centrifugal potential $\PhiC(r,\theta)$. As mentioned above, the latter must satisfy a cylindrical symmetry in order to preserve the barotropic relation. As a consequence, this approach cannot be used on potentials derived from shellular rotation profiles, \ie ~$\Omega = \Omega(r)$, for instance. It is worth noting, however, the large number of profiles that can be used, since any 1D rotation profile integrated using Eq.~\eqref{eq:def_PhiC} could lead to an eligible $\PhiC(s)$ in theory. In practice, it may be advisable to check that at least the Rayleigh criterion for stability is verified: $\partial_r(r^4\Omega^2) > 0$.

Once chosen, the integration of the rotation profile may lead to analytical expressions for the centrifugal potential, typical examples being those involving a solid rotation: 
\begin{equation}
    \Omega(s) = \Omega_0 \quad \rightarrow \quad \PhiC(r, \theta) = -\frac{1}{2}s^2\Omega_0^2
\end{equation} or a Lorentzian profile:
\begin{equation}
    \label{eq:lorentzian}
    \Omega(s) = \frac{1+\alpha}{1+\alpha (s/R_\mathrm{eq})^2}\Omega_0 \quad \rightarrow \quad \PhiC(r, \theta) = -\frac{1}{2}\frac{(1+\alpha)^2s^2\Omega_0^2}{1+\alpha (s/R_\mathrm{eq})^2},
\end{equation} where $s = r\sin\theta$ and $\Omega_0$ and $\alpha$ designate respectively the rotation rate on the equator and the relative difference on the rotation rate between the center and equator.

\subsection{Finding level surfaces and redistributing matter \label{par:level_surfaces}}

Once the total potential $\Phieff(r,\theta)$ has been calculated, level surfaces may be obtained by
finding isopotential lines.  As a first step, we see from Eq.~\eqref{eq:potential_1D} that the total potential just found must satisfy 
\begin{equation}
    \label{eq:potential_1D_rad}
    \Phieff(r,\theta) = \Phi_{\mathrm{sph}}(\rsph) + \mathrm{cnst},
\end{equation} meaning that the latter can be deduced from the initial (spherical)
potential to which a suitable constant has been added. The value of this constant is immediately found by applying the same relation at the model's centre:
\begin{equation}
    \label{eq:constant_value}
    \mathrm{cnst} = \PhiG(0, 0) + \PhiC(0, 0) - \Phi_{\mathrm{sph}}(0).
\end{equation}
Therefore, a given level surface $r_*(\theta)$ (corresponding to a certain $\zeta_*$) can be deduced from the freshly calculated potential $\Phieff(r,\theta)$ by satisfying Eq.~\eqref{eq:potential_1D_rad} for all $\theta$, using the constant provided by Eq.~\eqref{eq:constant_value}. As explained at the start of the section,
choosing level surfaces this way ensures that the correspondence between
$\rho(\zeta)$ and $\rho_{\mathrm{sph}}(\rsph)$ will be preserved for $\zeta=\rsph$.

Finding the level surfaces can be achieved in many different ways.  
In our algorithm, we decompose $f(r)=\Phieff(r,\theta)$ as a sum of Hermite splines since both $\Phieff$ and $\partial_r\Phieff$ are available.
Using the current level surfaces as a first guess, we can now use Newton's method to find the new $r$ values satisfying Eq.~\eqref{eq:potential_1D_rad}.
This approach can reach a high degree of accuracy fairly quickly given the efficiency of Newton's method and the fact that the positions of level surfaces change less and less during the successive iterations thus causing the first guess to become very close to the actual solution.

Once the new set of level surfaces, $r(\zeta,\theta)$, has been obtained,
redistributing the matter is trivial and corresponds to simply assigning the
$\rho$ values to the new surfaces. Note that, since all the above equations are solved in their rescaled form, a final step consists in updating the values involved in these scales, namely the mass, $M$, (as will be described in Sect.~\ref{sect:mass_growth}) and equatorial radius, $R_\mathrm{eq}$, of the model. This also implies rescaling the different non-dimensional variables, for instance:
\begin{eqnarray}
    \bar{\rho}_{i+1} &=& \bar{\rho}_{i} \frac{M_{i}}{M_{i+1}} \left(\frac{R_{i+1}}{R_{i}}\right)^3 \\
    \bar{P}_{i+1} &=& \bar{P}_{i} \left(\frac{M_{i}}{M_{i+1}}\right)^2 \left(\frac{R_{i+1}}{R_{i}}\right)^4
\end{eqnarray}
where $\bar{\rho}$ and $\bar{P}$ denote the dimensionless density and pressure profiles, and the indices $i$ and $i+1$ the iteration number. The algorithm then returns to the first step where the gravitational potential is calculated based on the matter distribution and iterations continue until the method converges. 

\subsection{Convergence \label{par:convergence}}

Convergence occurs once $r(\zeta,\theta)$ stops changing from one iteration to
the next, at least to within some given precision.  This can be measured in
various ways.  For the sake of simplicity, we check whether the variations
of $\Rpol/\Req$ has gone below a user-defined threshold in our algorithm.
In the rare cases where the algorithm fails to converge (typically at near-critical rotation rates), one can progressively increase the rotation rate with each iteration before reaching the nominal value.


\section{Specificities of this approach \label{sect:specificities}}

\subsection{Equation of state}

In the description provided in the previous section, one may note that, in contrast to various traditional and new SCF methods \citep{Jackson1970, Roxburgh2004, Jackson2005, MacGregor2007}, no mention to energy transfer has been made. In fact, by assuming that the effective relation between density and pressure is preserved, there is no need to address this question and the only equation explicitly solved in the program is Poisson's equation. As a direct consequence of this premise, the time needed to deform a given model is quite short, as shown in Tables~\ref{tab:performances_rad} \& \ref{tab:performances_sph}. It can seen that the computation time scales roughly with $NL$ in the spherical case and $NL^2$ in the spheroidal case.

\begin{table}[htbp]
\begin{center}
\caption{Performances (time | memory allocation) of RUBIS (spherical version) measured on a 1.9GHz Intel Core i7-8665U CPU with 4 cores (8 threads) processor. The model is a polytrope of index 3, rotating at $\Omega = 0.75\OmegaK$, $N$ being its radial resolution and $L$ the angular resolution of the 2D grid (as well as the number of spherical harmonics used). \label{tab:performances_rad}}~\\[-0.2cm]
\begin{tabular}{cccc}
\hline
\hline\\[-0.3cm]
\textbf{Spherical} & $N=1000$ & $N=2000$ & $N=4000$ \\
\hline \\[-0.2cm]
$L=25$ & 
$1.1$s \,|\, $0.3$GB &
$2.1$s \,|\, $0.4$GB &
$4.1$s \,|\, $0.8$GB \\
$L=51$ &
$2.3$s \,|\, $0.3$GB &
$4.0$s \,|\, $0.4$GB &
$7.0$s \,|\, $0.8$GB \\
$~~L=101$ & 
$4.0$s \,|\, $0.3$GB &
$7.2$s \,|\, $0.4$GB &
\!\!$12.4$s \,|\, $0.8$GB \\
\hline
\end{tabular}
\end{center}
\end{table}

\begin{table}[htbp]
\begin{center}
\caption{Same as Table~\ref{tab:performances_rad} but where Poisson's equation is solved in spheroidal coordinates. \label{tab:performances_sph}}~\\[-0.2cm]
\begin{tabular}{cccc}
\hline
\hline\\[-0.3cm]
\!\!\textbf{Spheroidal} & $N=1000$ & $N=2000$ & $N=4000$ \\
\hline \\[-0.2cm]
\!\!$L=25$ & 
$8.6$s \,|\, $0.5$GB &
\!\!$12.6$s \,|\, $0.6$GB &
\!\!$22.7$s \,|\, $0.8$GB \\
\!\!$L=51$ &
\!\!$23.7$s \,|\, $1.2$GB &
\!\!$35.4$s \,|\, $1.6$GB &
\!\!$61.8$s \,|\, $2.4$GB \\
\!\!$~~L=101$ & 
\!\!$76.8$s \,|\, $3.6$GB &
\!\!\!\!$114.1$s \,|\, $5.2$GB &
\!\!\!\!$181.0$s \,|\, $8.2$GB \\
\hline
\end{tabular}
\end{center}
\end{table}

Considering how drastic this simplification is, it is worth questioning its relevance and real meaning. Assuming that the model's structure satisfies a general equation of state depending on the temperature, $T$, and the chemical element abundances, $X_i$, $P(\rho, T, X_i)$ -- which need not be known in the present method --, and that matter is organised according to this equation in both the original and the deformed model, it can be seen that the conservation of the profile $\rho(P)$ along the level surfaces automatically implies a constraint on the thermal structure of the model. If, furthermore, the chemical composition is preserved during the transformation, this constraint merely results in the conservation of the temperature profile along the level surfaces. This consequence is somewhat approximate from an energetic point of view, since it has long been known that the thermal unbalance caused by rotation \citep{VonZeipel1924,Eddington1925} should give rise to effective temperature differences over the same level surface \citep{Zahn1992,Maeder1999}. In fact, any deformation assuming a conservative rotation profile will face this limitation. To try to estimate its impact on the structure, we propose in paragraph~\ref{par:ESTER} a comparison between the centrifugal deformation obtained with RUBIS and that obtained with a code including energy transfer, namely the ESTER code \citep{EspinosaLara2013, Rieutord2016}.

\subsection{Mass growth \label{sect:mass_growth}}

\begin{figure*}[htbp]
\begin{center}
\includegraphics[width=\textwidth]{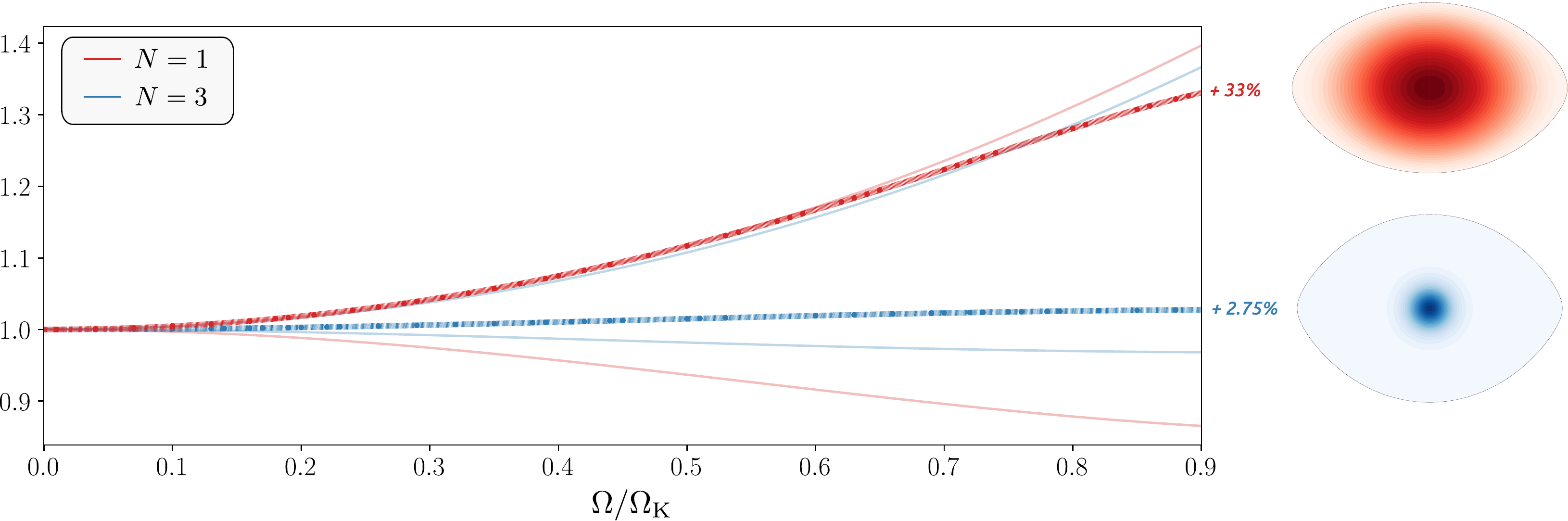}
\end{center}
\caption{Mass growth in polytropes of indices $N=1$ (red curves) and $N=3$ (blue curves) as a function of the normalised rotation rate in the case of a solid rotation. Upper and lower thin curves represent respectively the equatorial and polar radii in both polytropes while the thicker curves indicate their mass.  All quantities are expressed in units of the non-rotating models' parameters and the percentages on the right hand side give the relative mass increase at $\Omega = 0.9\OmegaK$ with $\OmegaK = \sqrt{\mathcal{G}M/R^3}$, the Keplerian rotation rate. The colour maps adjoining the curves depict the mass distribution in both models after deformation at $\Omega = 0.9\OmegaK$.
\label{fig:mass_poly}}
\end{figure*}

A direct consequence of preserving $\rho(P)$ is that the model mass is not conserved during the deformation. Indeed, although the density does not change on the level surfaces, the volume enclosed in each of them evolves during the deformation. This inevitably leads to a change in the total mass (and in most cases to an increase) as shown in Fig.~\ref{fig:mass_poly}. It should be emphasised that this is not an intrinsic inconsistency of the program but a consequence of the underlying assumptions; the deformation procedure we present must be seen as a purely mathematical transformation of the original model, not a dynamical one. In particular, the latter is not expected to preserve the total mass of the system as would be the case if a static model started to spin until it reached the desired rotation speed.

In addition to the rotation rate, the amount of mass growth also depends on the initial mass distribution of the model as illustrated in Fig.~\ref{fig:mass_poly}. Because the most deformed isopotentials are those located near the surface, models that concentrate most of the mass in their core (such as the polytrope of index $N=3$) will only see their mass change by a few percent, even at speeds close to the critical rotation rate. In contrast, more homogeneous models such as the $N=1$ polytrope will see isopotentials with considerable mass change significantly in volume, which can lead to a relative mass increase of over one third.

\subsection{Adaptive rotation rate \label{par:adaptive_rotation_rate}}

When going from one iteration to the next, one is faced with the following conundrum.
The centrifugal potential depends on the value of rotation rate.  
It then subsequently intervenes in the total potential and hence the calculation of new isopotential surfaces.  
In particular, this leads to a new determination of the equatorial radius (which is generally larger than the previous estimate).  
However, this radius is required for obtaining the rotation rate which intervenes in centrifugal potential, in order to ensure the ratio $\Omega/\OmegaK$ is preserved at the equator.  
Hence, there is an interdependence between the rotation rate and the equatorial radius.

\begin{figure}[htbp]
\begin{center}
\includegraphics[width=\columnwidth]{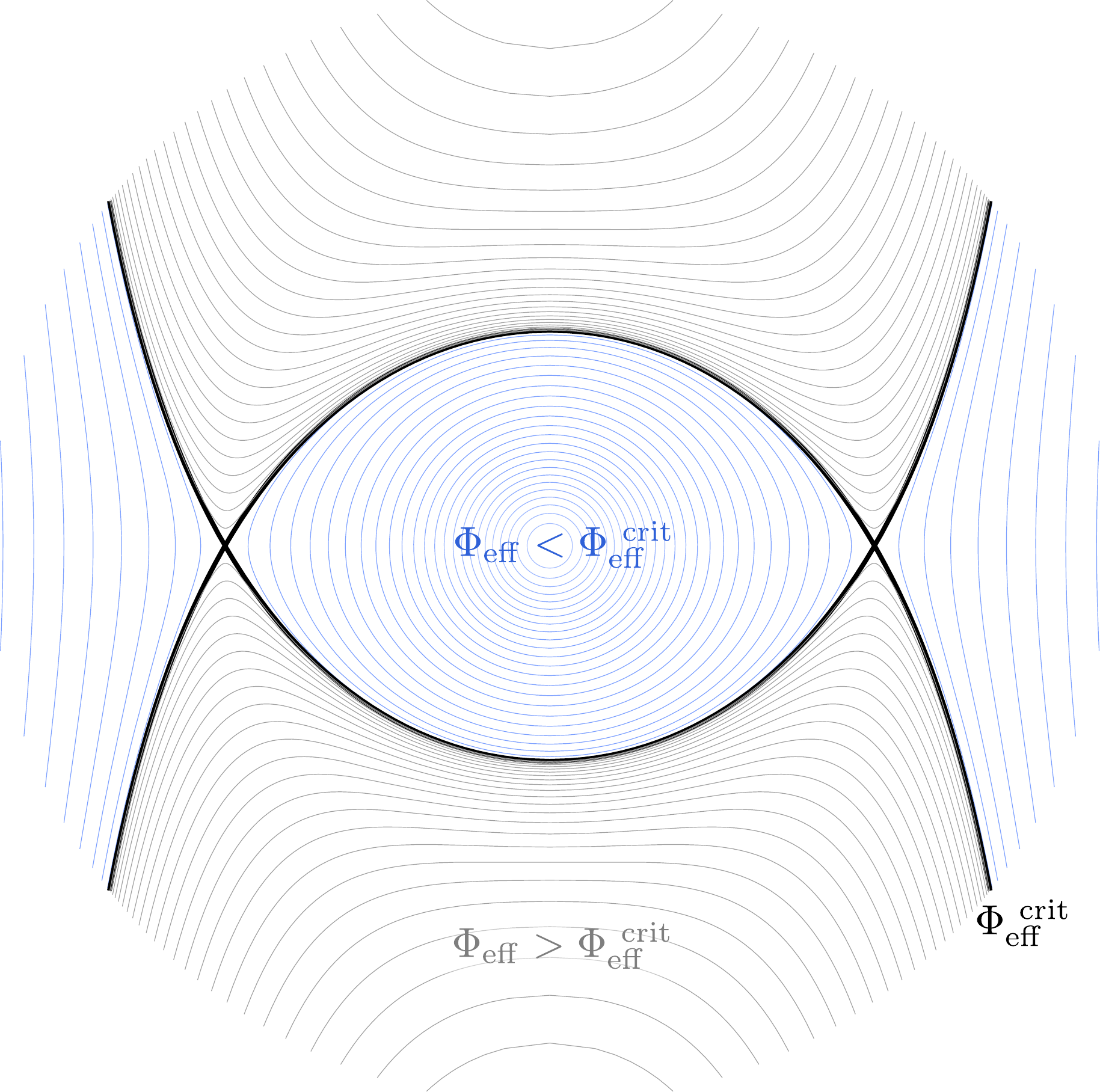}
\end{center}
\caption{Typical shapes of isopotentials in a meridional cross section, inside and outside a rotating model. The critical isopotential (denoted by its value, $\Phi_\mathrm{eff}^{~\mathrm{crit}}$) is shown in black, while level surfaces with higher and lower values respectively appear in grey and blue.
\label{fig:isopotential_scheme}}
\end{figure}

One may naively think that simply iterating the above algorithm, being a fixed-point scheme, will resolve this interdependence.
However, if the (dimensionless) target rotation rate is sufficiently close to critical, then one can easily exceed the critical rotation rate when the equatorial radius is updated.  
This then hampers calculating isopotential surfaces thus causing the iterations to stop prematurely.  
In order to resolve this interdependence, one needs to anticipate the value of the equatorial radius such that the target value of $\Omega/\OmegaK$ is reached at the equator.  
For the sake of clarity, we will now describe this using some equations.  
In what follows, the indices $i$ and $i+1$ will refer to the current and following iteration while the index $\infty$ will denote their limits. 
In order to avoid overloading an already cumbersome notation, dimensionless version of the quantities $R, \Omega, \Phi$ will be denoted as $r, \omega, \phi$, respectively.

From a mathematical point of view, the problem occurs when one has to find an isopotential with a value $\Phieff > \Phieff^{~\mathrm{crit}}$, the latter being defined as the value of the potential for which $\grad \Phieff \cdot \bm{e}_r = 0$ on the equator. 
Figure~\ref{fig:isopotential_scheme} provides a clear view of what happens in this case: since these isopotentials are not closed surfaces, Newton's method, as described in paragraph~\ref{par:level_surfaces}, cannot converge on the equator and the algorithm just breaks down. 
This issue might be surprising at first glance, since it is clear that even the value of  the highest isopotential (corresponding to the surface) should remain below $\Phieff^{~\mathrm{crit}}$ as long as the specified rotation profile, $\Omega(s)$, satisfies $\Omega_\mathrm{eq} / \OmegaK < 1$.

In a naive iterative scheme, however, it is actually possible to face this issue because the Keplerian break-up rotation rate, $\OmegaK^{~i} = \sqrt{\mathcal{G}M^i/{\Req^{~i}}^3}$, changes from one iteration to the next. 
Furthermore, since the equatorial radius increases faster than the mass as a function of the rotation rate (cf. Fig.~\ref{fig:mass_poly}), and since the dimensionless rotation rate on the equator $\omega_\mathrm{eq}$ does not change, the final equatorial rotation rate $\Omega_\mathrm{eq}^{~\infty} = \omega_\mathrm{eq}\OmegaK^{~\infty}$ will be smaller than the initial value, thus implying that at some point in the iterations it has decreased. 
In other words, the determination of the outermost isopotential with the current equatorial rotation rate, $\Omega_\mathrm{eq}^{~i}$, might have a value exceeding $\left(\Phieff^{~\mathrm{crit}}\right)^{~i+1}$ if $\Omega_\mathrm{eq}^{~i} / \OmegaK^{~i+1} > 1$. 
Naturally, the closer $\omega_\mathrm{eq}$ is to 1, the more likely this problem will occur.

To overcome this difficulty, one must find a new sequence of equatorial rotation rates, $\widetilde{\Omega}_\mathrm{eq}^{~i}$, such that the procedure ends up converging towards $\Omega_\mathrm{eq}^{~\infty} = \omega_\mathrm{eq}\OmegaK^{~\infty}$, without ever facing the criterion just stated above. 
The solution we found is to define this sequence as
\begin{equation}
    \widetilde{\Omega}_\mathrm{eq}^{~i} = \omega_\mathrm{eq}\widetilde{\Omega}_K^{~i}
\end{equation}where $\widetilde{\Omega}_K^{~i} = \sqrt{\mathcal{G}M^i/{(\Req^{~i+1})}^3}$. 
In terms of the current scaling, this solution boils down to adapting the dimensionless rotation rate during the iterations since $\widetilde{\Omega}_\mathrm{eq}^{~i}$ can be re-expressed as :
\begin{equation}
    \label{eq:adaptive_rotation_scaled}
    \widetilde{\Omega}_\mathrm{eq}^{~i} = \widetilde{\omega}_\mathrm{eq}^{~i}\OmegaK^{~i}
\end{equation} with 
\begin{equation}
    \label{eq:adaptive_rotation}
    \widetilde{\omega}_\mathrm{eq}^{~i} = \omega_\mathrm{eq} \frac{\widetilde{\Omega}_{\mathrm{K}}^{~i}}{\OmegaK^{~i}} = \omega_\mathrm{eq} (r_\mathrm{eq}^{~i+1})^{-3/2},
\end{equation} where $r_\mathrm{eq}^{~i+1}$ denotes the next equatorial radius expressed in the current scaling, $\Req^{~i+1}/\Req^{~i}$. 
Because the mass tends to increase from one iteration to the next, one can easily check that $\widetilde{\Omega}_K^{~i} < \OmegaK^{i+1}$ and therefore that the sequence we define will verify:
\begin{equation}
    \label{eq:prop_stability}
    \frac{\widetilde{\Omega}_\mathrm{eq}^{~i}}{\OmegaK^{~i+1}} = \widetilde{\omega}_\mathrm{eq}^{~i}\frac{\OmegaK^{~i}}{\OmegaK^{i+1}} = \omega_\mathrm{eq} \frac{\widetilde{\Omega}_K^{~i}}{\OmegaK^{~i}}\frac{\OmegaK^{~i}}{\OmegaK^{i+1}} < \omega_\mathrm{eq} < 1
\end{equation} for a given iteration $i$.
Moreover, as long as the sequence of equatorial radii converges towards a finite limit $\Req^{~\infty}$, the ratio between successive radii converge towards, $r_\mathrm{eq}^{~\infty} = 1$, thus proving with Eq.~\eqref{eq:adaptive_rotation} that the adaptive rotation rate we defined asymptotically approaches the user-specified value:
\begin{equation}
    \label{eq:prop_convergence}
    \lim_{i \rightarrow \infty} \widetilde{\omega}_\mathrm{eq}^{~i} = \omega_\mathrm{eq}.
\end{equation}

While this new definition seems to have all the desired properties (cf. Eqs.~\eqref{eq:prop_stability} \& \eqref{eq:prop_convergence}), it must be noted that the latter requires the evaluation of $\Req^{~i+1}$ at iteration $i$. 
Although anticipating the next equatorial radius is generally not possible in such an iterative procedure, an interesting feature in RUBIS enables its exact calculation. 
Since the effective potential only varies by an additive constant from one iteration to the next (see Eq.~\eqref{eq:potential_1D}), it may be pointed out that the difference 
\begin{equation}
    \begin{split}
        \delta\Phi &\equiv \Phieff(\zeta=1) - \Phieff(\zeta=0) \\
        &= \PhiG(R_\mathrm{eq}, \pi/2) + \PhiC(R_\mathrm{eq}, \pi/2) - \PhiG(0, \pi/2)
    \end{split}
\end{equation} does not change. 
Here it can be noted that the constant $\delta\Phi$ is known from the first iteration, after having solved Poisson's equation. 
Let us now place ourselves at iteration $i$ and try to anticipate the content of this equation at iteration $i+1$. 
Scaled by the current reference potential $\mathcal{G}M^i/R_\mathrm{eq}^{~i}$, the above equation becomes: 
\begin{equation}
    \label{eq:delta_phi_i}
    \delta\phi^i = \phi_g(r_\mathrm{eq}^{~i+1}, \pi/2) - \phi_g(0, \pi/2) - \int_0^{r_\mathrm{eq}^{~i+1}} x {\left[\Omega^i(x\Req^{~i})/\OmegaK^{~i}\right]}^2\, dx 
\end{equation} where we expressed the centrifugal potential using Eq.~\eqref{eq:def_PhiC} and defined $\delta\phi^i = \delta\Phi \times \left(\mathcal{G}M^i/R_\mathrm{eq}^{~i}\right)^{-1}$. 
In this equation, the scaled rotation profile $\Omega^i(s)$ depends on the iteration since its equatorial value changes with $i$ (cf. Eq.~\eqref{eq:adaptive_rotation_scaled}). 
The way the whole profile changes with its equatorial value in RUBIS can simply be described with the following scaling relation:
\begin{equation}
    \Omega^i(s) = \widetilde{\Omega}_\mathrm{eq}^{~i} \times \overline{\omega}(s/\Req^{~i})
\end{equation} where $\overline{\omega}$ designates the dimensionless profile such that $\overline{\omega}(1) = 1$. 
Injecting this expression into Eq.~\eqref{eq:delta_phi_i} and replacing $\widetilde{\Omega}_\mathrm{eq}^{~i}/\OmegaK^{~i}$ using Eq.~\eqref{eq:adaptive_rotation}, one obtains:
\begin{equation}
    \delta\phi^i = \phi_g(r_\mathrm{eq}^{~i+1}, \pi/2) - \phi_g(0, \pi/2) - \frac{\omega_\mathrm{eq}^{~2}}{{(r_\mathrm{eq}^{~i+1})}^3}\int_0^{r_\mathrm{eq}^{~i+1}} x \overline{\omega}^2(x)\, dx.
\end{equation}

Rescaling the $x$ variable in the integral so that it varies between 0 and 1, we get the final relation:
\begin{equation}
    \label{eq:delta_phi_i_final}
    \delta\phi^i = \phi_g(r_\mathrm{eq}^{~i+1}, \pi/2) - \phi_g(0, \pi/2) - \frac{\mathcal{R}\omega_\mathrm{eq}^{~2}}{r_\mathrm{eq}^{~i+1}}
\end{equation} with $\mathcal{R}$ a constant fixed by the choice of the rotation profile:
\begin{equation}
    \mathcal{R} = \int_0^1 x \overline{\omega}^2(x)\, dx.
\end{equation}

For instance, the choice of a uniform rotation profile leads to $\overline{\omega}(x) = 1$ and thus $\mathcal{R} = 1/2$, while choosing a Lorentzian profile (see Eq.~\eqref{eq:lorentzian}) yields:
\begin{equation}
    \overline{\omega}(x) = \frac{1+\alpha}{1+\alpha x^2} \quad \Rightarrow \quad \mathcal{R} = \frac{1+\alpha}{2}.
\end{equation}

We see that relation~\eqref{eq:delta_phi_i_final} is merely an equation of the type $f(r_\mathrm{eq}^{~i+1}) = \delta\phi^i$ which can be solved numerically for $r_\mathrm{eq}^{~i+1}$ using Newton's method and noticing that : 
\begin{equation}
    \frac{df}{dr} =  \left.\frac{\partial\phi_g}{\partial r}\right|_{\theta=\pi/2} + \mathcal{R}\,{\left(\frac{\omega_\mathrm{eq}}{r}\right)}^2
\end{equation} is known. 
Once $r_\mathrm{eq}^{~i+1}$ is found, the adaptive rotation rate follows immediately from Eq.~\eqref{eq:adaptive_rotation}.

\begin{figure*}[htbp]
\begin{center}
\includegraphics[width=\textwidth]{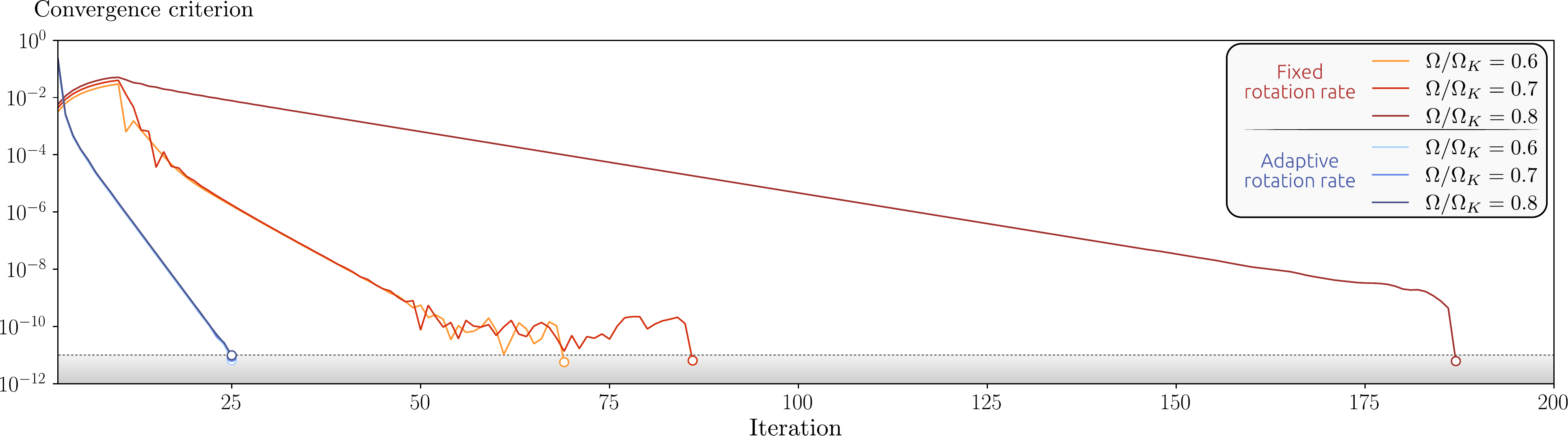}
\end{center}
\caption{Number of iterations needed to reach the convergence criterion (in this case $\left|(\Rpol/\Req)_{i+1} - (\Rpol/\Req)_{i}\right| < 10^{-11}$) for both the fixed rotation rate approach (red tone curves) and ``adaptive'' method (blue tone curves) presented in paragraph~\ref{par:adaptive_rotation_rate} (the radial and angular resolutions chosen were respectively $N=1000$ and $L=101$). 
For each approach, the degree of convergence as a function of iteration number is shown for three rotation rates. 
In the fixed-rotation approach, a transient phase in which the rotation speed gradually increases must be included at the beginning. 
In the case of the adaptive method, the three curves overlap.
\label{fig:convergence}}
\end{figure*}

In practice, this modification has a negligible numerical cost but offers a substantial gain in performance, both in terms of stability and convergence speed. 
Figure~\ref{fig:convergence} quantifies the benefits of this modification to the program by comparing the number of iterations required for convergence in both the fixed and adaptive rotation rate approaches. 
While the first method requires more and more iterations as $\Omega$ increases and ends up not converging beyond $0.8\OmegaK$, the adaptive approach reaches the desired criterion in a quasi-constant number of steps (about 25) whatever the rotation speed.

\begin{figure*}[htbp]
\begin{center}
\includegraphics[width=\textwidth]{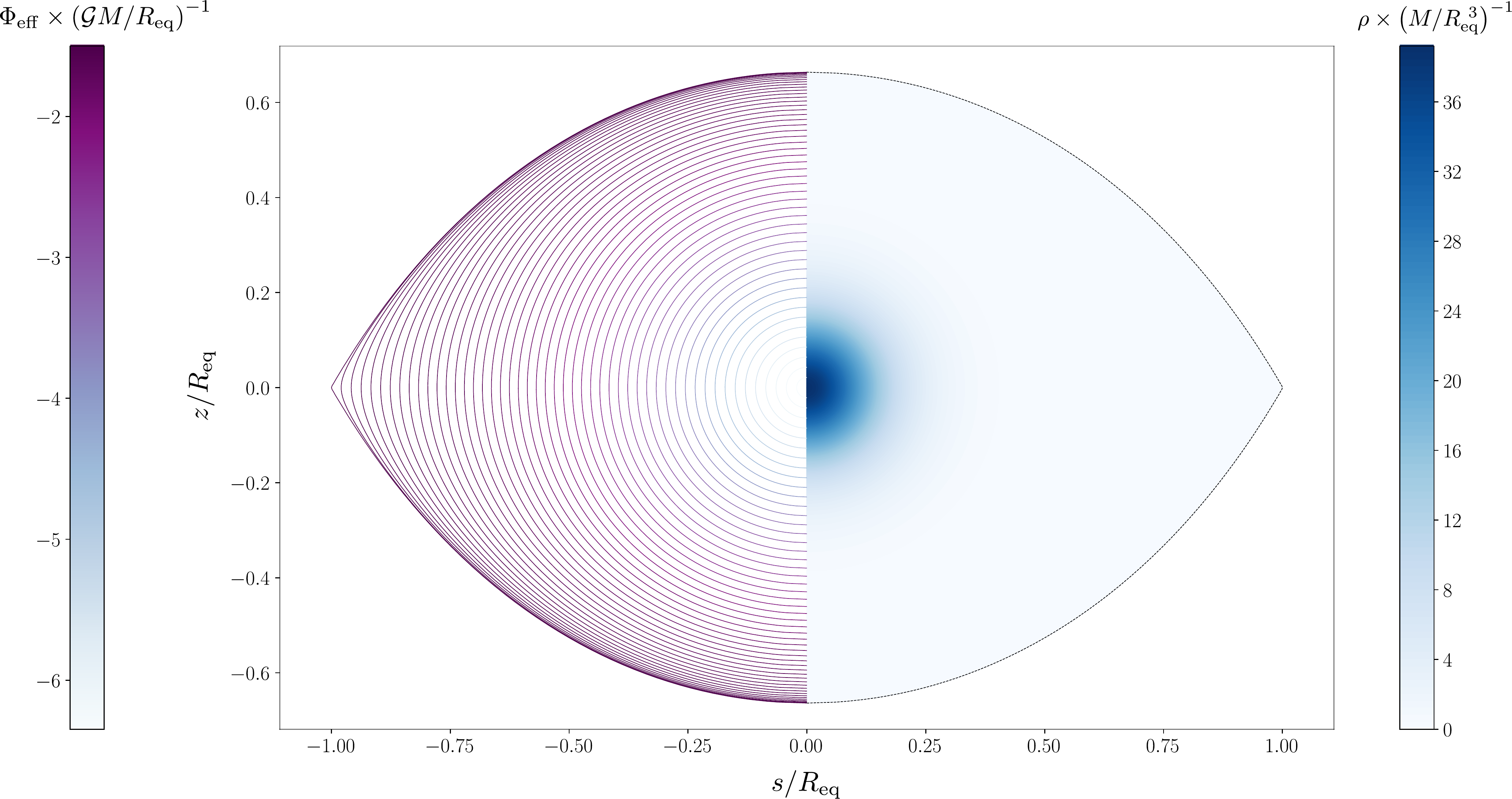}
\end{center}
\caption{Deformation of an $N=3$ polytrope at 99.99\% of the critical rotation rate. 
The left-hand side of the figure shows the shape of the level surfaces and their colour reflects the value of the effective potential. 
The right-hand side shows the mass distribution in the model.
\label{fig:0_9999_deform}}
\end{figure*}

In terms of stability, it is possible to reach speeds extremely close to the critical rotation rate. 
Figure~\ref{fig:0_9999_deform} shows the cross-section of an $N=3$ polytrope at $\Omega = 0.9999\OmegaK$. 
At such a speed, the last isopotential approaches the saddle point very closely thus leading to a well-defined cusp at the equator. 
It should be added that a high truncation order is required ($L = 500$) in order to properly resolve this region.

\section{Tests \label{sect:tests}}

As was pointed out in previous sections,  the method developed here preserves the polytropic relation between $\rho$ and $P$.  We will hereafter compare the deformations and structures obtained using the present method, and in some cases the resultant pulsation modes, with those generated by independent methods, namely the approaches developed in \citet{Rieutord2005}, in the ESTER code \citep{EspinosaLara2013, Rieutord2016}, and in the Concentric MacLaurin Spheroids (CMS) method \citep{Hubbard2012, Hubbard2013}.

\subsection{Polytropes}

\begin{figure*}[htbp]
\begin{center}
\includegraphics[width=\textwidth-4cm]{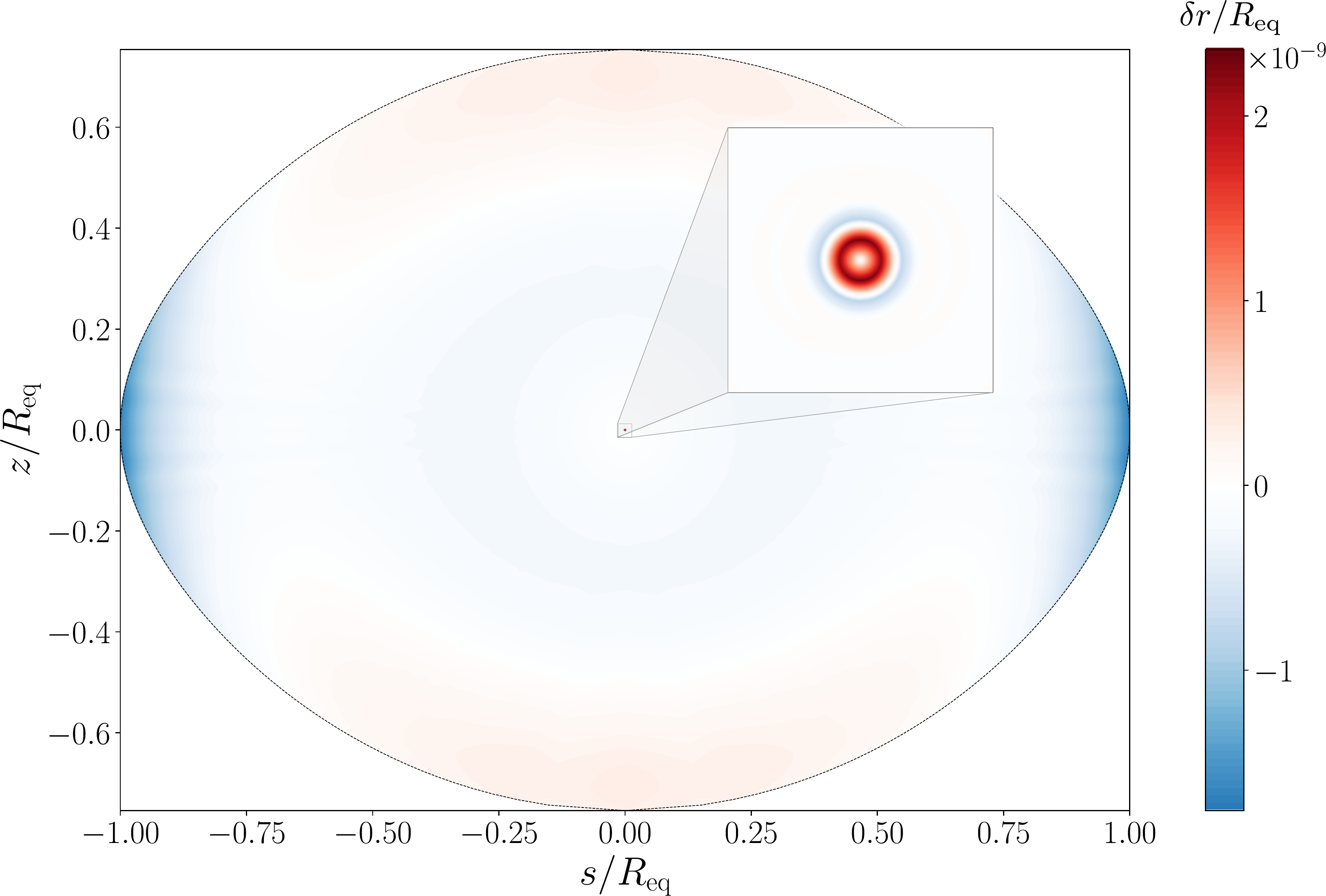}
\end{center}
\caption{Differences on level surfaces of the $0.8\,\OmegaK$ polytropic models.
Once more, the mappings have been normalised by the equatorial radius. The inset centred on the origin, where the differences are the greatest, is only 0.01 $R_\mathrm{eq}$ wide. 
\label{fig:polytrope_levels_diff}}
\end{figure*}

A particularly relevant -- and well-studied -- class of models to test are
polytropes. Indeed, polytropes with indices of $N=1.5$ and $3$ have
been as a first approximations of convective and radiative regions in stars
\citep[\eg][]{Eddington1926, Chandrasekhar1939}, while an index of $1$ has been used to model the envelope of planetary models such as Jupiter \citep[\eg][]{Stevenson1982}. Furthermore, the pulsation modes of such models have been calculated to a great accuracy by various authors \citep{Christensen-Dalsgaard1994, Lignieres2006, Reese2006, Ballot2010} and may be used as reference to test new methods.  In addition to this, rotating polytropes preserve their barotropic relation, so that the central assumption of our method is exactly verified for these particular models. Therefore, very small differences are expected compared to other deformation methods.

The polytropes generated using the present method have a
uniform radial grid with 1001 points and a colatitude grid of 101 points
distributed along a Gauss-Legendre collocation grid.  The models generated using
the \citet{Rieutord2005} method involves spectral methods in both the radial and
latitudinal directions.  A resolution of 81 points has been used in the radial
direction, and 51 spherical harmonics (with even $\ell$ values ranging from $0$
to $100$) for the horizontal structure. 

In order to account for the differences between the two methods, we chose to compare the positions of the level surfaces in a fast rotating $N=3$ polytrope. More specifically, we considered a uniform rotation at rate of $0.8 \OmegaK$.
This meant interpolating the
\citet{Rieutord2005} model to find the 1001 corresponding level surfaces as it is initially calculated using a mapping based on \citet{Bonazzola1998}. 
Figure~\ref{fig:polytrope_levels_diff} shows the result of this comparison, \ie the value of the differences
\begin{equation}
    \delta r(\zeta, \theta) = r^\mathrm{R05}(\zeta, \theta) - r^\mathrm{RUBIS}(\zeta, \theta)
\end{equation} inside the deformed models.
The maximum differences are of the order of $10^{-9}$ -- and, in most of the model, well below this value -- thus confirming the excellent agreement between the two methods. The most central (and pronounced) differences here are the result of the solving method used for Poisson's equation. In RUBIS, this equation is solved on $r^2$ rather than $r$ for regularity purposes, which may explain the $10^{-9}$ variations on $\PhiG$ in this region although the deformation is very small in practice.

\begin{figure}[htbp]
\begin{center}
\includegraphics[width=\columnwidth]{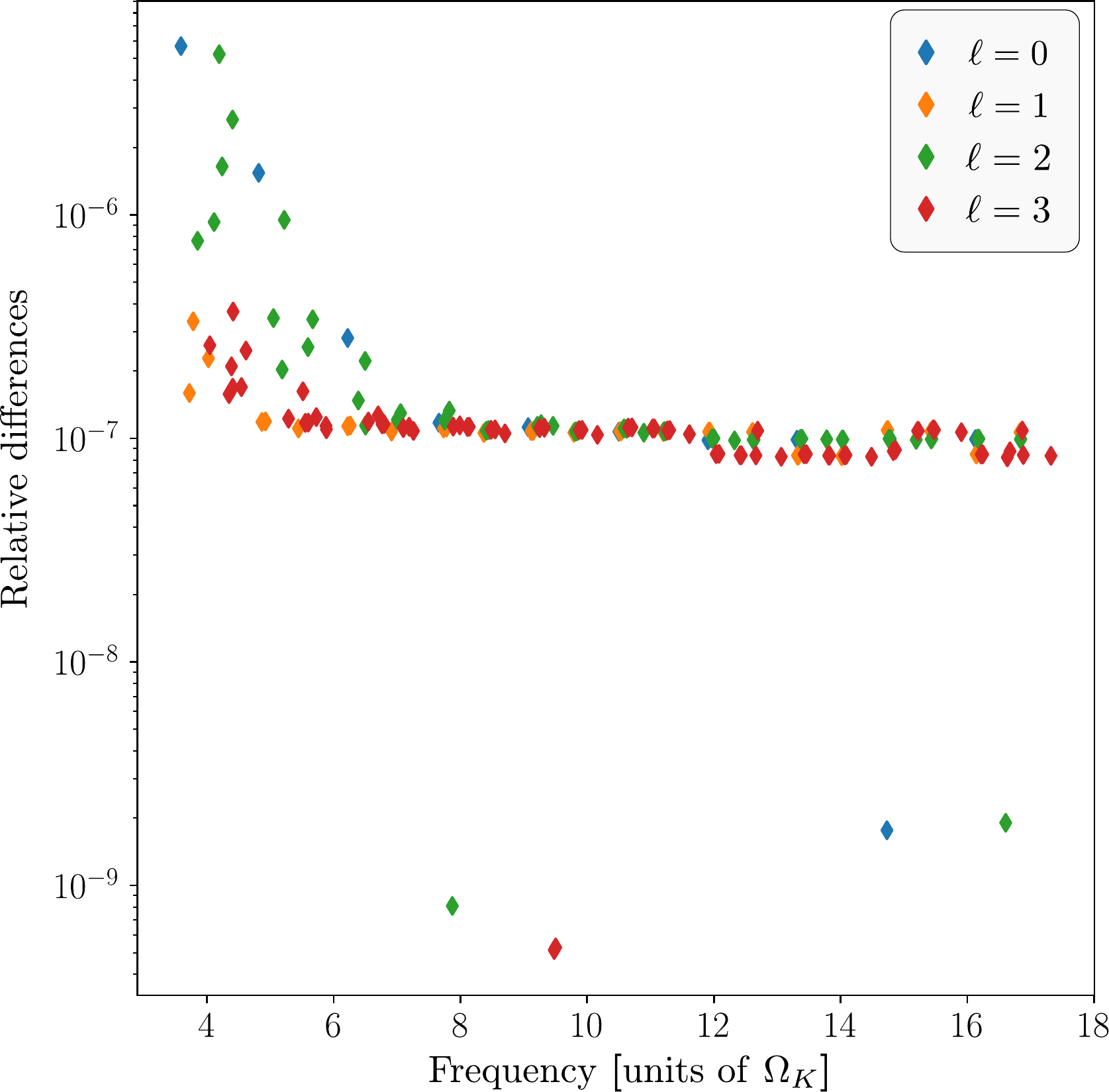}
\end{center}
\caption{Relative frequency differences between pulsation calculations in the
$N=3$, $\Omega=0.58946223\,\OmegaK$ polytrope from \citet{Reese2006} and an
equivalent model from the present method as a function of the oscillation frequency (expressed in units of the Keplerian rotation rate, $\OmegaK$). The different colours indicate the harmonic degree of the oscillation modes, obtained by correspondence with the non-rotating case. All azimuthal orders ($|m| \leq \ell$) are provided for each degree.
\label{fig:poly_freq_diff}}
\end{figure}

One of the goals of the method presented here is to produce models that may be
used for accurate pulsation calculations. We therefore compare the pulsation
frequencies of the most rapidly rotating model from \citet{Reese2006}, \ie\ an
$N=3$ polytrope rotating at $\Omega=0.58946223\,\OmegaK$ generated using
the \citet{Rieutord2005} method, with those of an equivalent model produced with
the present method.  This time we increased the radial resolution of the model
with the present method to $n=2000$ points using an unevenly spaced grid.  The
successive $\zeta$ positions (or $r$ positions of the precursor 1D model) are
given by:
\begin{equation}
\zeta_i=\sin\left(\frac{(i-1)\pi}{2(n-1)}\right), \qquad 1\leq i\leq n
\end{equation}
Such a grid has a roughly uniform spacing of $\sim \frac{\pi}{2n}$ around
$\zeta=0$ and a dense spacing that scales as $1/n^2$ near $\zeta=1$ thus making
it suitable for p-mode calculations. Figure~\ref{fig:poly_freq_diff} show the
relative frequency differences between the two models.  These differences turn
out to be of the order of $10^{-7}$ apart from the lowest frequency modes, thus
confirming the present method is fully able to produce accurate models suitable
for seismic calculations.

\subsection{ESTER models \label{par:ESTER}}

While RUBIS has been shown to reliably reproduce the structure of 2D polytropic models, its underlying assumptions suggest that the same cannot be said for more realistic structures. This will be evaluated in this section. In particular, a comparison will be made between structures derived from RUBIS and those derived from the ESTER code, which satisfy an energy balance in addition to the hydrostatic one. However, before this, let us first make use of ESTER to verify the relevance of RUBIS' central assumption, namely the conservation of the barotropic relation during the deformation process. To this end, we want to compare the relation between density and pressure in a rotating star obtained with ESTER and its non-rotating equivalent. However, there is an important aspect to consider in order to do this accurately. It was mentioned earlier that a deformation that preserves the barotropic relation does not conserve the mass of the model for the simple reason that its volume increases. Thus, it is to be expected that the density values present in models of same mass with and without rotation are not directly comparable. More precisely, the non-rotating model should be denser because of its smaller volume. This apparent problem can readily be solved: rather than comparing a rotating model with a static model of the \textit{same mass}, one simply needs to compare it with the model \textit{that will have the same mass after a deformation that preserves} $\rho(P)$, that is, deformed with RUBIS. Here, the difficulty arises from the impossibility of imposing the same rotation profile, RUBIS being limited to conservative rotation profiles whereas the ESTER profiles are fully differential (\ie\ non-conservative). To at least partly account for the change in mass, we compared an ESTER model with a model that reaches the same mass after having been deformed using a uniform rotation profile with $\Omega = \Omega_\mathrm{eq}^\mathrm{ESTER}$, where $\Omega_\mathrm{eq}^\mathrm{ESTER}$ is the equatorial rotation rate from ESTER. Most of the deformation should be taken into account in this way, although the differences $\delta\Omega$ between the two rotation profiles are of course part of the limitations to be kept in mind. This is confirmed by the relatively small difference in flattening between the two models (see Table~\ref{tab:ESTER_vs_RUBIS}).  Nonetheless, larger differences remain on the polar and equatorial radii, thus affecting the Keplerian break-up rotation rate as also shown in Table~\ref{tab:ESTER_vs_RUBIS}.

\begin{table}[htbp]
\begin{center}
\caption{Differences between the 2D ESTER and RUBIS models for various global properties. \label{tab:ESTER_vs_RUBIS}}~\\[-0.2cm]
\begin{tabular}{ccccc}
\hline
\hline\\[-0.3cm]
\textbf{Model} & $\Req$ & $\Rpol$ & $\Rpol/\Req$ & $\OmegaK$ \\
              & ($10^{11}$ cm)    & ($10^{11}$ cm) &  &  ($10^{-6}$ rad.s$^{-1}$) \\
\hline \\[-0.2cm]
ESTER & $1.459$ & $1.100$ & 0.7537 & 292.4 \\
RUBIS & $1.437$ & $1.087$ & 0.7566 & 299.3 \\
\hline
\end{tabular}
\end{center}
\end{table}

\begin{figure}[htbp]
\begin{center}
\includegraphics[width=\columnwidth]{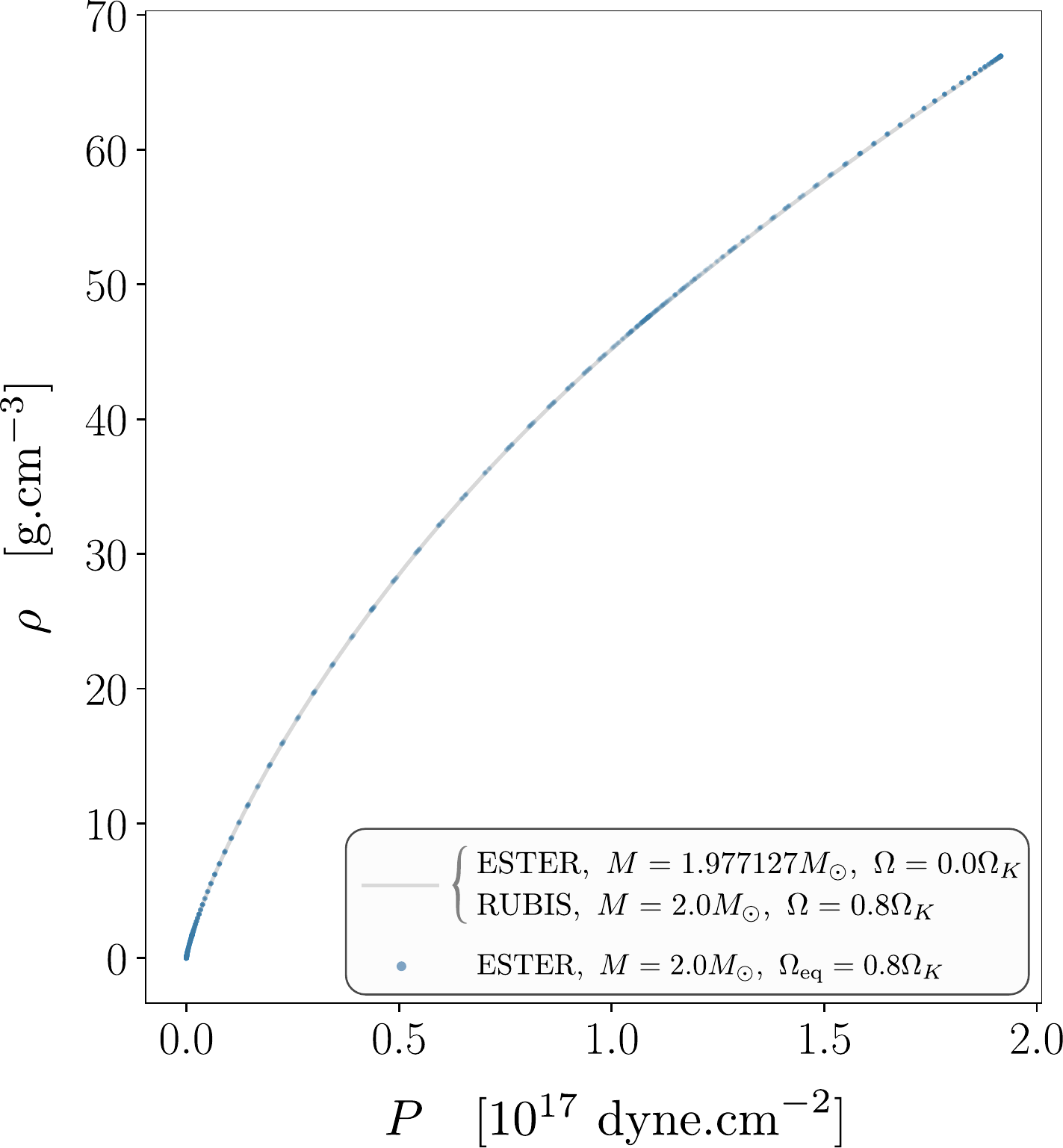}
\end{center}
\caption{Comparison between the 1D relationship $\rho(P)$ in a $1.977127M_\odot$ star without rotation (grey curve) and the couples $(\rho, P)$ from a $2M_\odot$ star obtained by ESTER using a differential rotation profile, the equatorial speed of which is $\Omega_\mathrm{eq} = 0.8\OmegaK$ (blue dots). The brackets indicate that the relation $\rho(P)$ shown here for the 1D model also corresponds to the one in the model deformed by RUBIS using the uniform rotation speed $\Omega = 0.8\OmegaK$ (the mass of the model then reaches $2M_\odot$).
\label{fig:rho_vs_P}}
\end{figure}

In Fig.~\ref{fig:rho_vs_P}, we compare the pairs ($\rho$, $P$) from a $2\,M_{\odot}$ ESTER model with a differential rotation verifying $\Omega_\mathrm{eq}^\mathrm{ESTER} = 0.8\OmegaK$ with the relation $\rho(P)$ from a 1D model of mass $1.977127\,M_\odot$.  When deformed by RUBIS using a uniform rotation profile with $\Omega = 0.8\OmegaK$, the latter leads to a $2\,M_\odot$ model that, by construction, verifies the same relation between density and pressure. In Fig.~\ref{fig:rho_vs_P}, two major properties stand out. First, although there is no relationship in the functional sense between $\rho$ and $P$ in the model deformed by ESTER, all pairs ($\rho$, $P$) seem to align on the same curve.  A high zoom level is necessary to reveal some thickness in this point distribution, resulting from angular differences.  It thus highlights the relevance of assuming the existence of such a relationship even to approximate more realistic cases, where energy transfer is taken into account. Moreover, this relation does not seem to be just any relation: the ($\rho$, $P$) pairs almost perfectly overlap with the $\rho(P)$ relationship of the non-rotating model! This observation is central and, in fact, constitutes a major motivation for developing a code such as RUBIS. Naturally, a closer inspection of the two structures reveals some differences, as will now be discussed.

\begin{figure*}[htbp]
\begin{center}
\includegraphics[width=\textwidth]{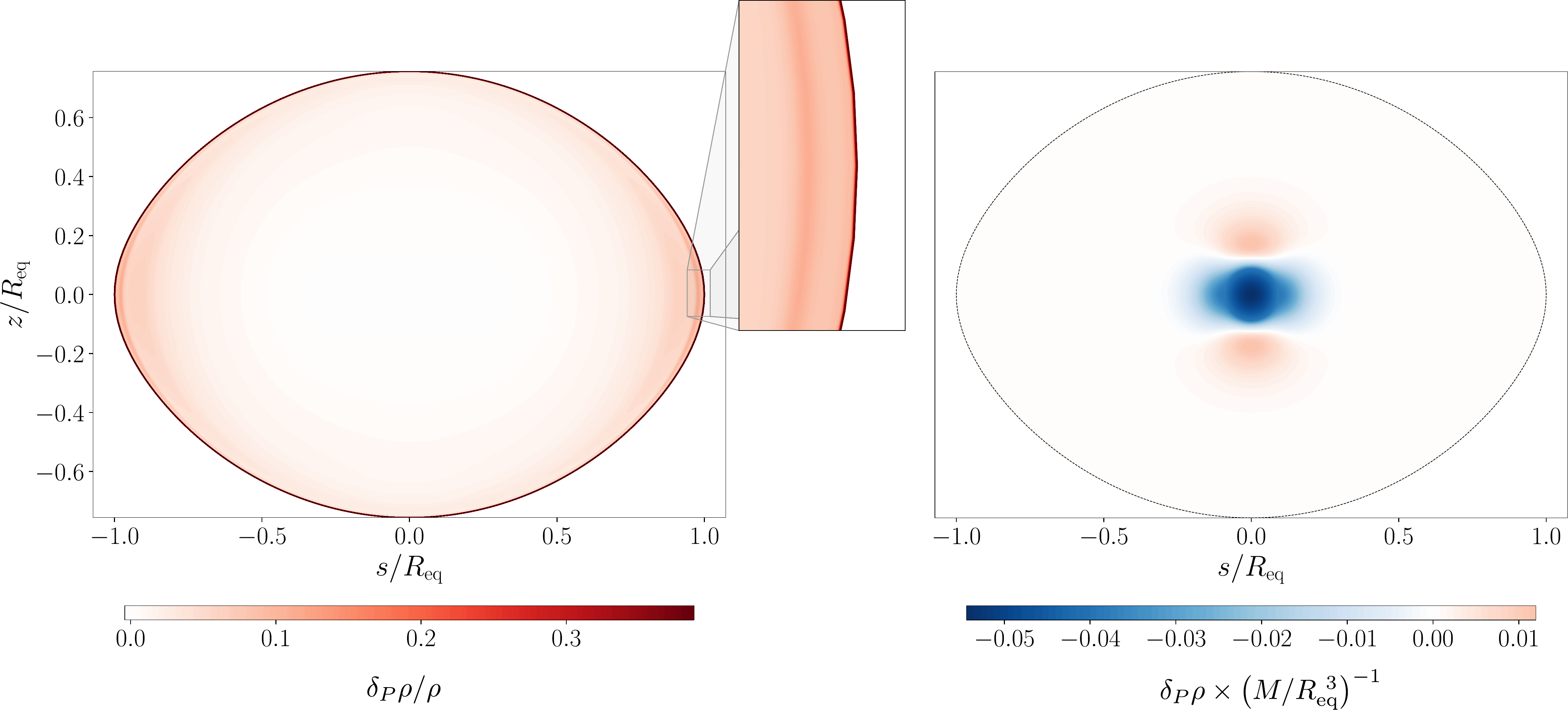}
\end{center}
\caption{\textit{Left panel:} relative density differences between the models computed with RUBIS and ESTER at given pressure values. The latter are placed according to the position where these values of pressure are met in the RUBIS deformation. The zoomed-in frame helps to reveal the differences in the most superficial layers at the equator. \textit{Right panel:} same as left panel but for the absolute differences in density (expressed this time in units of $M/R_\mathrm{eq}^{~3}$).
\label{fig:diff_ESTER}}
\end{figure*}

A drawback of Fig.~\ref{fig:rho_vs_P} is that it does not reveal in what regions of the star potential structural differences may arise.  In order to compare the structures, they need to be interpolated to the same pressure values\footnote{One may be tempted to interpolate the two models to the same isopotential lines, but we recall that such lines do not exist in ESTER models given its non-conservative rotation profile.}, the problem being that the methods used by ESTER and RUBIS do not use the same convention when defining the pseudo-radial variable, $\zeta$. For that, it is first necessary to interpolate the pressure as a function of $\zeta$ and $\theta$ in the ESTER model using the fact that it is defined on a multi-domain Gauss-Lobatto grid. We then find at which pairs $(\zeta^\mathrm{ESTER}, \theta)$ the pressure values in the RUBIS model corresponds, keeping in mind that the latter are defined at a fixed $\zeta^\mathrm{RUBIS}$. We now interpolate the ESTER density in order to evaluate it at $(\zeta^\mathrm{ESTER}, \theta)$ and we define:
\begin{equation}
    \delta_P \rho (\zeta^\mathrm{RUBIS}) = \rho(\zeta^\mathrm{RUBIS}) - \rho(\zeta^\mathrm{ESTER}, \theta).
\end{equation}
Here, the notation ``$\delta_P$'' indicates a difference at fixed pressure value, more specifically the one that is found at $\zeta^\mathrm{RUBIS}$. The quantity $\delta_P \rho$ therefore contains the deviations from barotropy obtained in the ESTER model, deviations that are possible to locate physically. We note that it also possible to define normalised differences, $\delta_P \rho / \rho$, which are represented, along with $\delta_P \rho$ in Fig.~\ref{fig:diff_ESTER}.

Regarding the normalised differences, the figure shows that they remain below 1\% in the innermost half of the star. However, they rapidly increase near the surface, eventually reaching 10\% in the ionisation region (on the equator), which is reflected in the red stripe visible in the zoomed-in frame. The largest relative differences exceed $1/3$ and are located in the very most superficial layers of the model (dark-red curve running around the star).

Absolute differences, on the other hand, offer an alternative picture. Near the centre, where we find near-solid rotation, the differences first take the form of a near-spherical function. Beyond a certain layer, however, these differences exhibit angular variations and reflect a more general differential rotation in the ESTER model. The absolute differences then tend rapidly towards 0.

These differences can mainly be attributed to two factors. First, as mentioned above, the two models do not have the same rotation profile. It might be possible that some of these differences can be taken into account by defining a ``best-fitting conservative profile'' as:
\begin{equation}
    \Omega(s) = \frac{\int_0^{Z(s)} \rho \Omega^\mathrm{ESTER}(z, s)\, dz}{\int_0^{Z(s)} \rho(z,s)\, dz}
\end{equation}
with $Z(s)$ the position of the surface at a distance $s$ from the rotation axis. However, this factor alone does not explain all of the observed differences and one can also expect that the thermal structure, verifying a more complex equilibrium in the ESTER model, is only approximately reproduced by a model deformed with RUBIS. Although the hydrostatic and thermal structures can be seen as uncorrelated as long as the equation of state is not specified -- in the sense that there are an infinite number of thermal structures and equations of state leading to the same equilibrium --, it is obviously not the case for a model aiming to verify an energy transfer equilibrium. One can therefore naturally expect such differences between RUBIS and ESTER.

In practice, both representations given in Fig.~\ref{fig:diff_ESTER} have their own relevance depending on the models usage. For example, the high relative differences on the surface can be expected to play an important role when calculating high degree pressure modes. On the other hand, gravity modes or global quantities sensitive to mass distribution such as gravitational moments may be more sensitive to the second representation.

\begin{figure}[htbp]
\begin{center}
\includegraphics[width=\columnwidth]{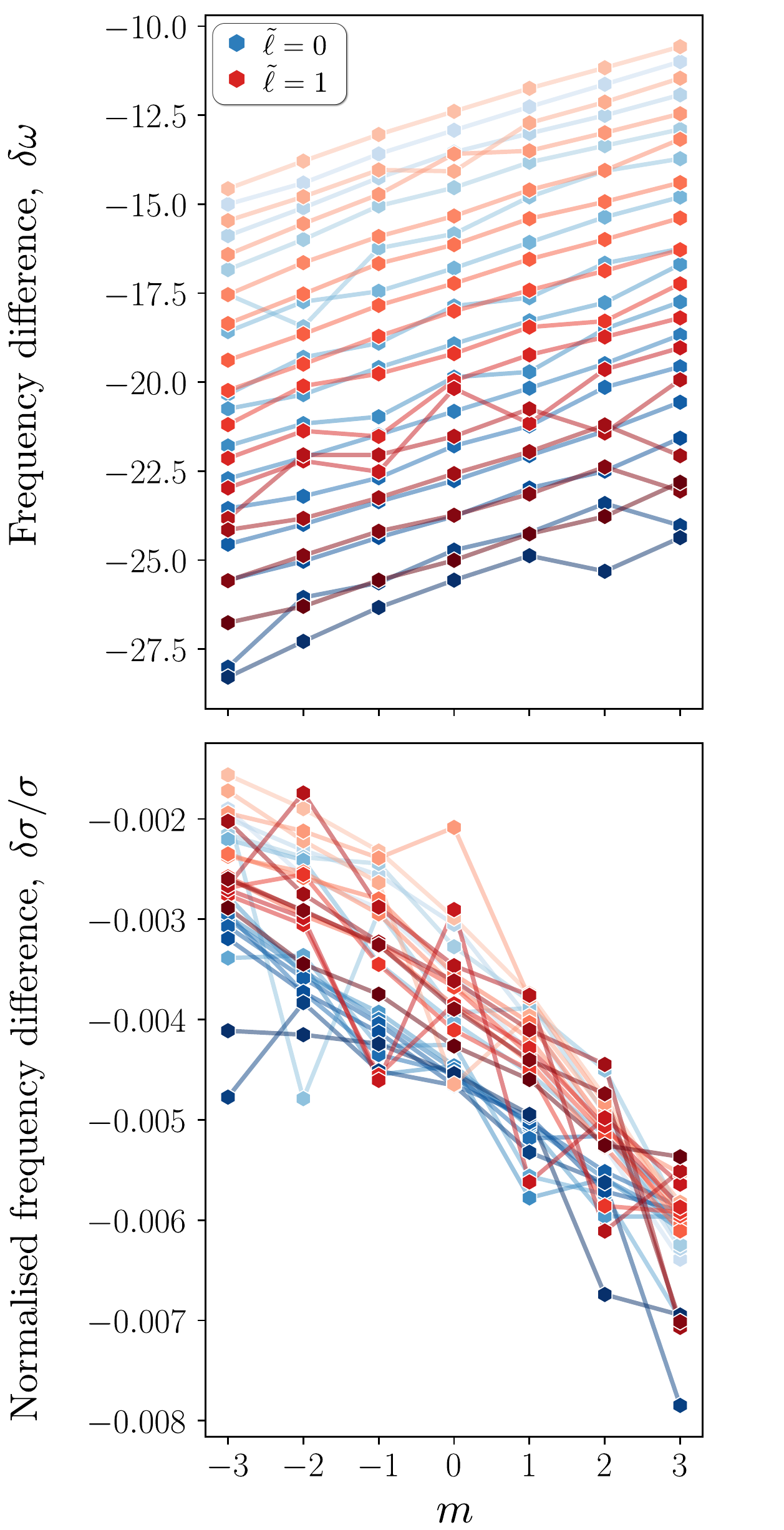}
\end{center}
\caption{Frequency differences in $\mu$Hz (upper panel), as defined in Eq.~\eqref{eq:freq_diff_scaled}, along with their normalised counterparts (lower panel) introduced in Eq.~\eqref{eq:freq_diff}. These differences are shown as a function of the azimuthal order, $m$, for the $\Tilde{\ell} = 0$ (blue colour shades) and $\Tilde{\ell} = 1$ (red colour shades) island modes. The colour shade indicates the mode's pseudo-radial order, $\Tilde{n}$  (higher orders correspond to darker colours).
\label{fig:freq_diff}}
\end{figure}

\begin{figure*}[htbp]
\begin{center}
\includegraphics[width=\textwidth]{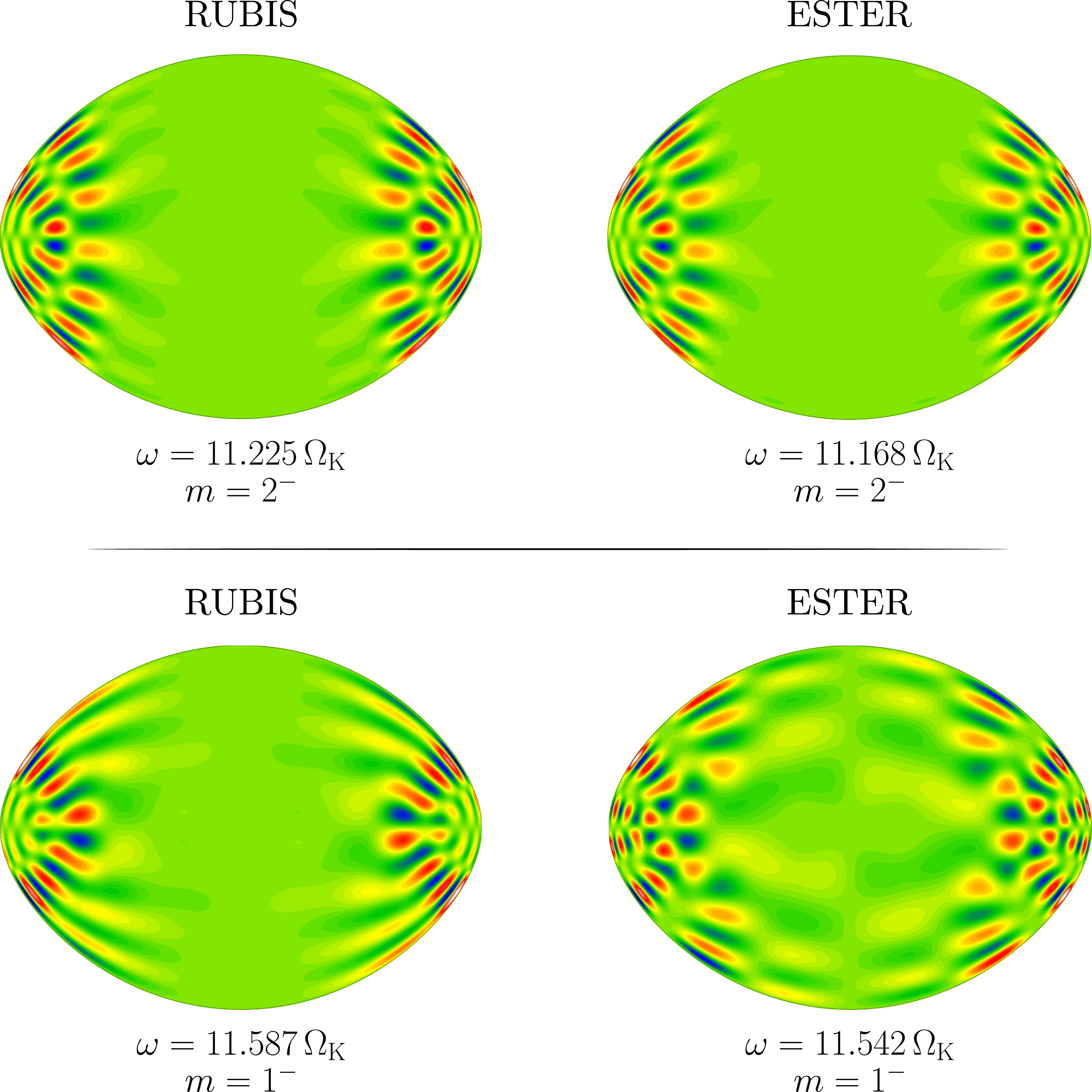}
\end{center}
\caption{Two oscillation modes (upper and lower parts of the figure) in both the RUBIS and ESTER models (resp. left and right sides). The mode on top is an $\Tilde{\ell} = 1$, $m = 2$ antisymmetric (with respect to the equator) mode whereas the one on the bottom corresponds to $\Tilde{\ell} = 0$, $m = 1$. The oscillation frequency of each mode is indicated in units of $\OmegaK$.
\label{fig:modes}}
\end{figure*}

In the following, we will account for the impact of those differences on the models' oscillation frequencies.  Using the Two-dimensional Oscillation Program (TOP) \citep{Reese2006,Reese2009a}, we computed and identified the oscillation modes corresponding to the $\Tilde{\ell} = 0, 1$, $\tilde{n}=9,23$ acoustic island modes. We recall that island modes are the rotating counterparts to low degree acoustic modes.  They focus around a period ray orbit which circumvents the equator.  The quantum number $\tilde{n}$ corresponds to the number of nodes along the orbit's path whereas $\Tilde{\ell}$ is the number of nodes parallel to the orbit (see \citealt{Lignieres2008}, \citealt{Reese2008b}, and \citealt{Pasek2012} for a more detailed definition of $\Tilde{\ell}$ and $\tilde{n}$ and their link with usual spherical quantum numbers). 
The frequency differences, $\delta\omega$, defined as:
\begin{equation}
    \label{eq:freq_diff_scaled}
    \delta \omega = \omega^\mathrm{ESTER} - \omega^\mathrm{RUBIS}
\end{equation} are represented in the upper panel of Fig.~\ref{fig:freq_diff} as a function of the azimuthal order $m$ for both the $\Tilde{\ell} = 0$ and $\Tilde{\ell} = 1$ oscillation modes. 

The first point to highlight is that the differences are not distributed around 0 but exhibit a clear negative offset. The reason for this is the different frequency scales resulting from the differences in radii (see Table~\ref{tab:ESTER_vs_RUBIS}).  This is confirmed by the fact that the differences increase with $\tilde{n}$, as indicated by the colour shades in Fig.~\ref{fig:freq_diff}.  It also explains the trend as a function of the azimuthal order $m$.  Indeed, in our convention, modes with negative $m$ values are prograde and therefore have higher frequencies.  This then highlights the difference in frequency scale.

If we now normalise the frequencies to account for these scaling effects, as well as for the order of magnitude of the frequency differences, we obtain the differences shown in the
lower panel of Fig.~\ref{fig:freq_diff}.  These differences may be expressed as follows:
\begin{equation}
    \label{eq:freq_diff}
    \frac{\delta \sigma}{\sigma} = \left[\frac{\omega^\mathrm{ESTER}}{\OmegaK^{~\mathrm{ESTER}}} - \frac{\omega^\mathrm{RUBIS}}{\OmegaK^{~\mathrm{RUBIS}}}\right]{\left(\frac{\omega^\mathrm{ESTER}}{\OmegaK^{~\mathrm{ESTER}}}\right)}^{-1}.
\end{equation}

The negative offset has been mostly removed, along with the previously observed trend with $\Tilde{n}$. This representation also reveals that the normalised frequency differences are of few thousandths. Naturally, the latter are considerably higher than in the case of a comparison between barotropic models with the same rotation profiles (see Fig.~\ref{fig:poly_freq_diff}).

It is also interesting to note that the trend as a function of $m$ has been reversed.  Indeed, by removing the scaling effects, this brings to light more subtle effects related to the rotation profiles of the two models as we will now explain. Following \cite{Reese2021}, the rotational splittings in ESTER can be expressed (neglecting the Coriolis force) as:
\begin{equation}
    \omega_{-m} - \omega_{+m} \simeq 2m\Omega_\mathrm{eff},
    \label{eq:Omega_eff}
\end{equation} where 
\begin{equation}
    \Omega_\mathrm{eff} = \int_V \Omega(r, \theta)\,\mathcal{K}(r, \theta)\,dV,
\end{equation} is a weighted average of $\Omega$, and $\mathcal{K}$ a (mode-dependent) rotation kernel defined as:
\begin{equation}
    \mathcal{K}(r, \theta) = \frac{1}{2}\left[\frac{\rho |\bm{\upxi}_{+m}|^2}{\displaystyle \int_V \rho |\bm{\upxi}_{+m}|^2\, dV} + \frac{\rho |\bm{\upxi}_{-m}|^2}{\displaystyle \int_V \rho |\bm{\upxi}_{-m}|^2\, dV}\right],
\end{equation}where $|\bm{\upxi}_{\pm m}|$ designates the retrograde ($+m$) (resp. prograde ($-m$)) mode amplitude.

Because the rotation profile is differential in the ESTER model, the value of $\Omega_\mathrm{eff}$ is likely to differ from $\Omega_\mathrm{eq}$. Moreover, because $\Omega(r, \theta)$ tends to be higher than $\Omega_\mathrm{eq}$ in the regions probed by $\mathcal{K}(r, \theta)$, it is to be expected that:
\begin{equation}
    \omega_{+m} - \omega_{-m} > 2m\Omega_\mathrm{eq},
\end{equation}
where we've made use of Eq.~(\ref{eq:Omega_eff}).  In contrast, the model deformed by RUBIS is rotating uniformly, thus leading to:
\begin{equation}
    \omega_{+m} - \omega_{-m} \simeq 2m\Omega_\mathrm{eq}.
\end{equation}
By comparing the non-dimensional version of the two above equations, and recalling that the non-dimensional equatorial rotation rate is the same in both models, we obtain:
\begin{equation}
    \delta(\sigma_{-m} - \sigma_m) \simeq 2m\delta\Omega_{\mathrm{eff}}/\OmegaK
\end{equation}
where
\begin{equation}
    \delta\Omega_\mathrm{eff} = \int_V (\Omega(r, \theta) - \Omega_\mathrm{eq})\,\mathcal{K}(r, \theta)\,dV > 0.
\end{equation}

Regarding the structure of the modes themselves, we compare in Fig.~\ref{fig:modes} the following oscillation modes obtained in the RUBIS and ESTER models: $(\Tilde{n}, \Tilde{l}, m)=(13, 1, 2)$ and $(13, 0, 1)$ (resp. upper and lower panels). The first comparison shows a typical example of a mode that is almost identical in the RUBIS and ESTER models while the second one exhibits clear differences in the two modes. More specifically the mode in the ESTER model is considerably altered by an avoided crossing while its impact is just beginning to emerge on its counterpart in the RUBIS model. Overall, the oscillation modes that possess well-defined structures are very similar, and even some avoided crossings are well-reproduced in both models.

\subsection{Model of Jupiter}

To illustrate the capabilities of RUBIS in deforming planetary models, we consider here the centrifugal deformation of a model of Jupiter. This specific case is very interesting in practice, be it for determining Jupiter's structure by fitting its gravitational moments \citep{Hubbard2012, Hubbard2013, Debras2018} or for interpreting oscillations modes obtained thanks to projects following the Jovian Oscillations through Velocity Images At several Longitudes (JOVIAL) project \citep{JOVIAL2019}.

\begin{figure}[htbp]
\begin{center}
\includegraphics[width=\columnwidth]{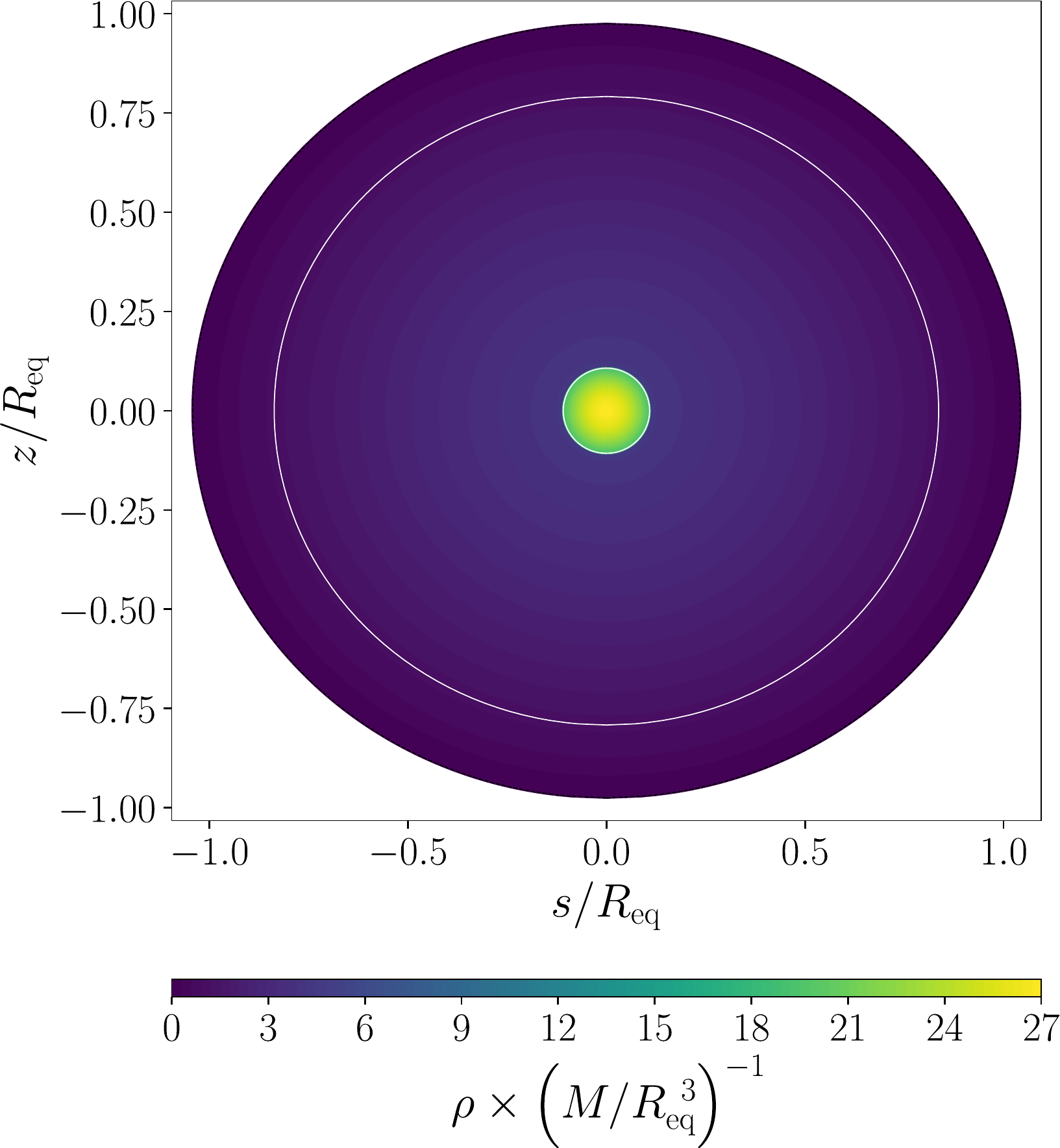}
\end{center}
\caption{Mass distribution in a Jovian model after imposing a uniform rotation at $\Omega \simeq 0.298656 \OmegaK$ using RUBIS. The white contours indicate the location of the model's discontinuities.
\label{fig:jupiter}}
\end{figure}

In order to illustrate RUBIS' capabilities in deforming planetary models, we have considered a Jupiter model provided by the Code d'Evolution Planetaire Adaptatif et Modulaire (CEPAM) \citep{Guillot1995}. This model presents a strong density discontinuity due to the presence of a solid core (responsible for a change of $\sim 75\%$ in density), making it an ideal application to test the program's stability. In Fig.~\ref{fig:jupiter}, we represent the mass distribution in the Jovian model when imposing a solid rotation rate of $\Omega \simeq 0.298656 \OmegaK$, which corresponds to the one used in \cite{Debras2018}. The most central discontinuity, which corresponds to the solid core, is clearly visible in the colour change. A more discrete discontinuity caused by the metallic to molecular phase change in the envelop has been highlighted with a white contour. Finally, it is worth noting that the code convergence subsists for Jovian models exceeding $0.9\,\OmegaK$, although their concrete applications become somewhat uncertain...

Another advantage of using RUBIS for deforming planetary models lies in the excellent accuracy it provides at a very low numerical cost. This will be illustrated here by confronting a classic problem of Jovian science: the calculation of gravitational moments able to satisfy the observational constraints of \textit{Juno}. Based on numerous flybys of the planet \citep{Bolton2017b}, the probe is able to provide very reliable estimates of Jupiter's gravitational moments (see Table~\ref{tab:moments_Jupiter}), defined as :
\begin{equation}
    J_\ell = -\frac{1}{MR_\mathrm{eq}^{~\ell}} \int_V r^\ell \rho(r, \theta) P_\ell(\cos\theta)\,dV,
\end{equation}where $P_\ell$ designates the $\ell$-th Legendre polynomial, and therefore to account for the departures from a spherically symmetric matter distribution caused by the rotation.

\begin{table}[htbp]
\begin{center}
\caption{Gravitational moments of Jupiter measured by Juno after 17 Jovian passes \citep{Durante2020}. \label{tab:moments_Jupiter}}~\\[-0.2cm]
\begin{tabular}{ccc}
\hline
\hline\\[-0.3cm]
 & Measured values & $3\sigma$ uncertainties \\
\hline \\[-0.2cm]
$J_2 \times 10^2$ & 
~~$1.46965735$ &
~~$0.00000017$ \\
$J_4 \times 10^4$ &
$-5.866085 ~~~~$  &
~~$0.000024 ~~~~$ \\
$J_6 \times 10^5$ & 
~~$3.42007 ~~~~~~$ & 
~~$0.00067 ~~~~~~$ \\
$J_8 \times 10^6$ &  
$-2.422 ~~~~~~~~~~$ &
~~$0.021 ~~~~~~~~~~$ \\
$J_{10} \times 10^7$ & 
~~$1.81 ~~~~~~~~~~~~$ &
~~$0.67 ~~~~~~~~~~~~$ \\
$J_{12} \times 10^8$ & 
~~$6.2 ~~~~~~~~~~~~~~$ & 
$19.0 ~~~~~~~~~~~~~~$ \\
\hline
\end{tabular}
\end{center}
\end{table}

\begin{table*}[htbp]
\begin{center}
\caption{Gravitational moments of the $N=1$ polytrope found by different methods, compared with the analytical values. Reds decimals indicate inaccurate digits while grey decimals indicate digits subject to numerical error (cf. text).} \label{tab:moments_poly}~\\[-0.2cm]
\begin{tabular}{cccccc}
\hline
\hline \\[-0.3cm]
 & Analytical values$^\star$ & ToF, $3^\mathrm{rd}$ order$^{~*}$ & CMS, $N_\mathrm{sph} = 512^{~\dagger}$ & CMS, $N_\mathrm{sph} = 512^{~\ddagger}$ & RUBIS \\
\hline \\[-0.25cm]
$q$ & 
$0.089195487$ & 
$0.089195487$ & 
$0.089195487$ & 
$0.089195487$ & 
$0.089195487$\\[0.05cm]
$J_2 \times 10^{2}$ & 
~~$1.3988511$ &
~~$1.39{\color{mdtRed}\bm{94099}}$ &
~~$1.398{\color{mdtRed}\bm{9253}}$ & 
~~$1.398{\color{mdtRed}\bm{9239}}$ &
~~$1.3988511$ \\
$J_4 \times 10^{4}$ &
$-5.3182810$  &
$-5.3{\color{mdtRed}\bm{871087}}$ & 
$-5.318{\color{mdtRed}\bm{7997}}$ &  
$-5.318{\color{mdtRed}\bm{7912}}$  &
$-5.3182810$ \\
$J_6 \times 10^{5}$ & 
~~$3.0118323$ & 
~~$3.{\color{mdtRed}\bm{9972442}}$ & 
~~$3.01{\color{mdtRed}\bm{22356}}$ & 
~~$3.01{\color{mdtRed}\bm{22298}}$ & 
~~$3.0118323$ \\
$J_8 \times 10^{6}$ &  
$-2.1321157$ &  
--- &
$-2.132{\color{mdtRed}\bm{4628}}$ &  
$-2.132{\color{mdtRed}\bm{4581}}$ &  
$-2.132115{\color{mdtGray}\bm{7}}$ \\
$J_{10} \times 10^{7}$ & 
~~$1.7406712$ & 
--- &
~~$1.740{\color{mdtRed}\bm{9925}}$ & 
--- &
~~$1.74067{\color{mdtGray}\bm{1}}{\color{mdtRed}\bm{1}}$ \\
$J_{12} \times 10^{8}$  & 
$-1.5682195$ & 
--- &
$-1.568{\color{mdtRed}\bm{5327}}$ &  
--- &
$-1.5682{\color{mdtGray}\bm{195}}$ \\
$J_{14} \times 10^{9}$ & 
~~$1.5180992$ & 
--- &
~~$1.518{\color{mdtRed}\bm{4156}}$ & 
--- &
~~$1.518{\color{mdtGray}\bm{09}}{\color{mdtRed}\bm{80}}$ \\
\hline
\end{tabular}
\end{center}
$^\star$ : \cite{Wisdom2016} \\
$^*$ : \cite{Hubbard1975,Zharkov1978} \\
$^\dagger$ : \cite{Hubbard2013} \\
$^\ddagger$ : \cite{Debras2018}
\end{table*}

However, a considerable difficulty in comparing these observational values with those of Jovian models is the accuracy that deformation codes can achieve for these moments. A classic benchmark for testing this is the calculation of the gravitational moments of the $N=1$ polytrope with deformation parameter $q = \omega^2 = 0.089195487$, for the reason that they can be calculated analytically. We provide, in Table~\ref{tab:moments_poly}, a comparison between these analytical values and the numerical estimates from several methods: the Theory of Figure (ToF) to order $q^3$, the CMS method using $512$ spheroids and the method presented here. The studies from which these values are taken can be found below the table. We facilitate the comparison between columns by indicating in red the digits that do not match the analytical values.

In order to reach an optimal accuracy, we used a radial grid of $10\,000$ points with a Gauss-Legendre grid of $101$ points in the angular direction for a maximum harmonic degree $L$ of $100$. Numerical errors on the moments were assessed via the variance of the results over more than 100 runs with slight variations of the radial grid. Digits that are correct in average but may change between runs are indicated in grey in Table~\ref{tab:moments_poly}.

The results are quite impressive. Whereas the ToF method leads to an absolute error of about $10^{-5} - 10^{-6}$, and the CMS method a relative error of $10^{-4}$, RUBIS exhibits an absolute error of about $10^{-13}$, pushed back to $10^{-14} - 10^{-16}$ when considering the average estimates. Moreover, this accuracy can be achieved at a fairly low numerical cost. For instance, the RUBIS deformation process described here was performed using a 1.9GHz Intel Core i7-8665U CPU with 4 cores (8 threads) processor, running in 33.9s on average and requiring 4.1 GB of memory.

Finally, it should be emphasised that, compared to other methods such as the Consistent Level Curves (CLC) method which can achieve arbitrarily high accuracy on the moments \citep{Wisdom2016}, RUBIS was designed to be able to take into account density discontinuities in a consistent manner. Its ability to overcome the difficulties faced by CMS when increasing the number of spheroids \citep{Debras2018} and thus guarantee very high accuracy even for the first moments makes it a reasonable choice when searching for Jovian models subject to the constraints of textit{Juno}.

\section{Conclusion \label{sect:conclusion}}

In this article we present RUBIS, a fully Python-based centrifugal deformation program which can be accessed from this \href{https://github.com/pierrehoudayer/RUBIS}{GitHub repository}. The program takes in an input 1D (spherically symmetric) model and returns its deformed counterpart by applying a conservative rotation profile specified by the user. More specifically, the code only needs the density profile as a function of radial distance, $\rho(r)$, from the reference model in addition to the surface pressure, $P_0$, in order to perform the deformation. The program is particularly lightweight thanks to the central assumption which consists in preserving the relation between density and pressure when going from the 1D to the 2D structure. The latter makes it possible, in particular, to avoid the standard complications arising from energy conservation in the resulting model \citep{Jackson1970, Roxburgh2004, Jackson2005, MacGregor2007}. In this sense, the method is analogous to the one presented by \cite{Roxburgh2006}, but simpler since it does not require the calculation of the first adiabatic exponent, $\Gamma_1$, during the deformation process, thus bypassing the need to explicitly specify an equation of state.

As a result, the only equation solved, in practice, by the program is Poisson's equation, $\Delta \PhiG = 4\pi\mathcal{G}\rho$, thereby leading to very fast computation times, even when high angular accuracy is required, thus making it a potentially valuable tool for multidimensional evolutionary applications. Another feature of the method is its excellent stability, which enables the deformation of models at speeds very close to the critical rotation rate. Finally, the code has been designed to allow both stellar and planetary models to be deformed, by dealing with potential discontinuities in the density profile. This is made possible by solving Poisson's equation in spheroidal rather than spherical coordinates whenever a discontinuity is present.

By design, RUBIS is able to reach a high degree of accuracy when deforming polytropic structures, making them well suited to the calculation of oscillation frequencies. We have also shown that the centrifugal deformations derived from RUBIS keep a certain degree of relevance even when approximating more complex baroclinic structures, and this even when they exhibit a differential rather than conservative rotation profile. Concerning their frequencies themselves, although the differences with respect to calculations in more realistic models are significant from the observational point of view, the frequencies from RUBIS models nonetheless are accurate to a few tenths of a percent once scaling effects have been taking into account.  This is useful for providing a first estimate of stellar parameters when interpreting observed pulsation spectra which can then guide more costly searches using more realistic models. Finally, we also demonstrated the ability of the program to deform discontinuous structures such as planetary models and illustrated its accuracy when calculating gravitational moments. The results are promising compared to the existing alternatives and highlight RUBIS' viability as a model adjustment tool for fitting the measurements coming from the \textit{Juno} probe.

\begin{acknowledgements}
The authors thank the referee for their comprehensive reading and very detailed comments which significantly contributed to making the article clearer.
PSH and DRR warmly thank Pr. Tristan Guillot for providing them with a model of Jupiter which is deformed for illustrative purposes in Fig.~\ref{fig:jupiter}. PSH and DRR acknowledge the support of the French Agence Nationale de la Recherche (ANR) to the ESRR project under grant ANR-16-CE31-0007 as well as financial support from the Programme National de Physique Stellaire (PNPS) of the CNRS/INSU co-funded by the CEA and the CNES.
DRR also acknowledges the support of the ANR to the MASSIF project under grant ANR-21-CE31-0018-01.
\end{acknowledgements}

\bibliographystyle{aa}
\bibliography{biblio}

\appendix

\section{Poisson's equation and interface conditions derivation in spheroidal coordinates \label{app:Poisson}}

Here we briefly retrace the derivation of Eq.~\eqref{eq:Poisson_spheroidal_proj} as well as the boundary conditions~\eqref{eq:interface_phi} \& \eqref{eq:interface_dphi}, while clarifying the notations associated with the coupling integrals $\mathcal{P}^{\ell\ell '}_{\cdot\cdot}$. First of all, Eq.~\eqref{eq:Poisson_spheroidal} can quickly be retrieved using the fact that $\Delta \PhiG = \bm{\nabla}\cdot\bm{\nabla}\PhiG$, which in curvilinear coordinates is written as (using Einstein's summation convention):
\begin{equation}
    \Delta \PhiG = \frac{\partial_i\left(\sqrt{|g|}\, \partial^i\PhiG\right)}{\sqrt{|g|}} = \frac{\partial_i\left(g^{ij}\sqrt{|g|}\,\partial_j\PhiG\right)}{\sqrt{|g|}}.
\end{equation}
Here, $g^{ij}$ denotes the covariant components of the metric tensor and $g$ its determinant. Since $\partial_\varphi\PhiG = 0$, the only components needed are: $g^{\zeta\zeta} = (r^2+\rt^2)(r^2\rz^2)^{-1}$, $g^{\zeta\theta} = g^{\theta\zeta} = -\rt(r^2\rz)^{-1}$ and $g^{\theta\theta} = r^{-2}$, in addition to $g = r^4\rz^2\sin^2\theta$. Multiplying by $r^2\rz$, the sum then results in four terms:
\begin{equation}
    \begin{split}
        r^2\rz \Delta\PhiG =~& \dz\left(\frac{r^2+\rt^2}{\rz}\dz\PhiG\right) - \dz\left(\rt\dt\PhiG\right) \\
        &- \frac{1}{\sin\theta}\dt\left(\rt\sin\theta \dz\PhiG\right)
        + \frac{1}{\sin\theta}\dt\left(\rz\sin\theta \dt\PhiG\right).
    \end{split}
\end{equation}

Expanding the last three terms and replacing $\Delta \PhiG$ by its actual value according to Poisson's equation, $4\pi\mathcal{G}\rho$, leads to:
\begin{equation}
    \dz\left(\frac{r^2+\rt^2}{\rz}\dz\PhiG\right) - 2\rt\dzt\PhiG - \Delta_\mathcal{S}r\,\dz\PhiG + \rz\Delta_\mathcal{S}\PhiG = 4\pi\mathcal{G}r^2\rz\rho,
\end{equation}
with $\Delta_\mathcal{S} \equiv \sin^{-1}\theta \dt\left(\sin\theta \dt \right) = \dtt+ \cot\theta \dt$. We recognise Eq.~\eqref{eq:Poisson_spheroidal}. Decomposing the gravitational potential over the (normal) Legendre polynomials:
\begin{equation}
    \PhiG(\zeta, \theta) = \sum_{\ell'=0}^\infty \PhiG^{\ell'}(\zeta) P_{\ell'}(\cos\theta),
\end{equation}
we now obtain:
\begin{equation}
    \begin{split}
        \sum_{\ell'=0}^\infty &\dz\left(\frac{r^2+\rt^2}{\rz}P_{\ell'}\dz\PhiG^{\ell'}\right) \\
        & - \left[2\rt \dt P_{\ell'} + \Delta_\mathcal{S}r\,P_{\ell'}\right]\dz\PhiG^{\ell'} \\
        & - \ell'(\ell'+1) \rz P_{\ell'}\PhiG^{\ell'} \qquad\qquad = 4\pi\mathcal{G}r^2\rz\rho
    \end{split}
\end{equation}
using the fact that $\Delta_\mathcal{S}P_{\ell} = -\ell(\ell+1)P_{\ell}$. 

Since $\int_{-1}^{-1} P_\ell(\mu) P_{\ell'}(\mu)\,d\mu = \delta_{\ell\ell'}$, one can project this equation on the $\ell$-th polynomial which leads to:
\begin{equation}
    \sum_{\ell' = 0}^L \dz\left(\mathcal{P}_{\zeta \zeta}^{\ell\ell'} \dz \PhiG^{\ell'}\right) - \mathcal{P}_{\zeta \theta}^{\ell\ell'}\dz \PhiG^{\ell'} - \mathcal{P}_{\theta \theta}^{\ell\ell'} \PhiG^{\ell'} = 4 \pi \mathcal{G} {(r^2\rz)}_\ell~ \rho
\end{equation}
by introducing the coupling integrals:
\begin{align}
    \mathcal{P}_{\zeta \zeta}^{\ell\ell'} &= \int_{-1}^{1} \frac{r^2+\rt^2}{\rz}P_\ell P_{\ell'}\,d(\cos\theta), \\
    \mathcal{P}_{\zeta \theta}^{\ell\ell'} &= \int_{-1}^{1}\left[2\rt P_\ell \dt P_{\ell'} + \Delta_\mathcal{S}r\,P_\ell P_{\ell'}\right]\,d(\cos\theta), \\
    \mathcal{P}_{\theta \theta}^{\ell\ell'} &= \ell'(\ell'+1) \int_{-1}^{1} \rz P_\ell P_{\ell'}\,d(\cos\theta)
\end{align}
and the following $r^2\rz$ decomposition:
\begin{equation}
    r^2\rz = \sum_{\ell'=0}^\infty (r^2\rz)_{\ell'} P_{\ell'}(\cos\theta),
\end{equation}
thus proving Eq.\eqref{eq:Poisson_spheroidal_proj}. It must noted that the spectral decomposition of $\rho$ does not appear since it only depends on $\zeta$.

Now looking at the boundary conditions to impose on the domain interfaces, the two quantities that need to be continuous are $\PhiG$ and its gradient.  The continuity of $\PhiG$ leads to the first condition on $\PhiGl$, \ie\ its continuity:
\begin{equation}
    \label{eq:interface_phi}
    {\left(\PhiGl\right)}^- = {\left(\PhiGl\right)}^+,
\end{equation}
where `-` and `+` denote quantities below and above the interface. In order to find a second boundary condition, the gradient of $\PhiG$ must first be re-expressed from the natural basis ($\bm{\mathrm{b}}^\zeta$, $\bm{\mathrm{b}}^\theta$) to an orthogonal basis such as the spherical one ($\hat{\bm{e}}_r$, $\hat{\bm{e}}_\theta$):
\begin{equation}
    \begin{split}
        \bm{\nabla}\PhiG &= \dz\PhiG \, \bm{\mathrm{b}}^\zeta + \dt\PhiG \, \bm{\mathrm{b}}^\theta \\
        &= \frac{\dz\PhiG}{\rz} \hat{\bm{e}}_r + \frac{1}{r}\left(\dt\PhiG - \frac{\rt}{\rz}\dz\PhiG\right) \hat{\bm{e}}_\theta,
    \end{split}
\end{equation} by using the relations:
\begin{align}
    \bm{\mathrm{b}}^\zeta &= \frac{1}{r_\zeta} \hat{\bm{e}}_r  - \frac{r_\theta}{r r_\zeta} \hat{\bm{e}}_\theta,\\
    \bm{\mathrm{b}}^\theta &=  \frac{1}{r}\hat{\bm{e}}_\theta.
\end{align}

We see that, in order for this gradient to be continuous, both $\dz\PhiG/\rz$ and $\dt\PhiG - (\rt/\rz)\dz\PhiG$ must be preserved across the interface. At this point, it can be noted that the interface must necessarily follow an isobar for the pressure gradients to compensate on both sides. Since isobars and isopotentials coincide, this surface corresponds to a constant value of $\zeta$ between $0$ and $1$, denoted $\zeta_*$. Moreover, both $r$ (for obvious reasons) and $\PhiG$ (from the first boundary condition) are continuous through this surface. Therefore, evaluating the $\theta$ derivative at $\zeta_*$ on both sides leads to:
\begin{align}
    \rt^-(\zeta_*, \theta) &= \rt^+(\zeta_*, \theta) \\
    \dt\PhiG^-(\zeta_*, \theta) &= \dt\PhiG^+(\zeta_*, \theta)
\end{align}
Therefore, both $\rt$ and $\dt\PhiG$ are continuous, and preserving $\dt\PhiG - (\rt/\rz)\dz\PhiG$ is equivalent to preserving $\dz\PhiG/\rz$. Let us now express this condition on the $\PhiGl$. We have:
\begin{equation}
    \frac{\dz\PhiG}{\rz} = \sum_{\ell' = 0}^\infty \frac{\dz\PhiG^{\ell'}}{\rz} P_{\ell'}
\end{equation} and projecting this decomposition on the $\ell$-th polynomial leads to the second boundary condition:
\begin{equation}
    \label{eq:interface_dphi}
    \sum_{\ell' = 0}^\infty \left(\mathcal{P}^{\ell\ell'}_\mathrm{BC}\right)^- \left(\dz\PhiG^{\ell'}\right)^- = \sum_{\ell' = 0}^\infty \left(\mathcal{P}^{\ell\ell'}_\mathrm{BC}\right)^+ \left(\dz\PhiG^{\ell'}\right)^+
\end{equation}with:
\begin{equation}
    \mathcal{P}_\mathrm{BC}^{\ell\ell'} = \int_{-1}^{1} \frac{1}{\rz} P_\ell P_{\ell'}\,d(\cos\theta).
\end{equation}

\end{document}